\documentstyle[12pt,epsf,epsfig,graphicx]{article}
\newcommand{\ba}{\begin{array}}
\newcommand{\ea}{\end{array}}
\newcommand{\be}{\begin{equation}}
\newcommand{\ee}{\end{equation}}
\newcommand{\bea}{\begin{eqnarray}}
\newcommand{\eea}{\end{eqnarray}}
\newcommand{\beq}{\begin{equation}}
\newcommand{\eeq}{\end{equation}}

\def\GSW_sign{}

\def\bra{\langle}
\def\ket{\rangle}
\def\a{\alpha}
\def\b{\beta}
\def\g{\gamma}

\def\e{\epsilon}

\def\l{\lambda}
\def\m{\mu}
\def\n{\nu}
\def\G{\Gamma}

\def\to{\rightarrow}

\def\bra{\langle}
\def\ket{\rangle}
\def\a{\alpha}
\def\b{\beta}
\def\g{\gamma}

\def\e{\epsilon}

\def\p{\pi}

\def\l{\lambda}
\def\m{\mu}
\def\n{\nu}
\def\G{\Gamma}

\def\mub{\mu_b}

\def\to{\rightarrow}

\def\bra{\langle}
\def\ket{\rangle}
\def\a{\alpha}
\def\b{\beta}
\def\g{\gamma}

\def\e{\epsilon}

\def\p{\pi}

\def\l{\lambda}
\def\m{\mu}
\def\n{\nu}
\def\G{\Gamma}

\def\mub{\mu_b}

\def\to{\rightarrow}

\def\o{\over}
\def\b{\begin{equation}}
\def\e{\end{equation}}

\def\kpn{K^+\rightarrow\pi^+\nu\bar\nu}

\def\klpn{K_L\rightarrow\pi^0\nu\bar\nu}

\begin{document}
 
\thispagestyle{empty}
\begin{flushright}
CERN-TH/2001-146\\
April 2001
\end{flushright}

\vspace*{1.5cm}
\centerline{\Large\bf Inclusive Rare B Decays}
\vspace*{2cm}
\centerline{{\bf Tobias Hurth}}
\bigskip
\centerline{\it Theory Division, CERN, CH-1211 Geneva 23,
                Switzerland}
 
\vspace*{1.5cm}
\centerline{\bf Abstract}
We review the present status of rare $B$ decays, focusing on 
inclusive decay modes and their role in our search for 
new physics. 
We also briefly discuss 
direct CP violation in rare $B$ decays and  the
 rare kaon decays
$K^+\rightarrow \pi^+\nu\bar{\nu}$ and 
\mbox{$K_L \rightarrow \pi^0 \nu \bar{\nu}$}, 
which offer complementary opportunities for  
precision flavour physics. 
\vspace*{0.3cm}
\noindent

\vspace*{7cm}
{Based 
on invited talks given at 
the 8th International Conference on Supersymmetries in Physics
(\,{\bf    SUSY2K}\,), 26 June - 1 July,\, 2000,\, CERN,\, Geneva,\, 
Switzerland,
and at the 5th International Symposium on Radiative Corrections
({\bf \mbox{RADCOR} 2000}\,), 11 - 15 September,\, 2000,\, Carmel,\, California.}

\vfill

\newpage
\pagenumbering{arabic}

\setcounter{equation}{0}
\section{Introduction}
\label{intro}

Flavour physics deals with that  part of the standard model (SM) which 
distinguishes between the three  generations of fundamental fermions.
It is still a mystery why there are exactly three generations.
Also the origin of the fermion masses  and their mixing is unknown; 
in particular, the SM does not explain the hierarchical  pattern of these 
parameters. Flavour physics can be regarded as the least tested part of the 
SM.  This is reflected in the rather large error bars of 
several flavour parameters such as the mixing parameters 
at the $20 \%$ level \cite{Parodi}, which has to be compared with 
errors smaller than $1 \%$ in high energy electroweak 
precision experiments.

\begin{figure}
\begin{center}
\epsfig{figure=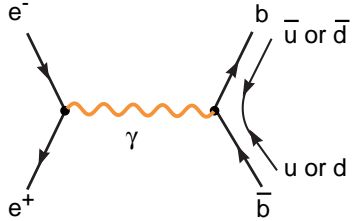,width=5cm}
\end{center}
\vspace{-1cm}
\caption{$e^+ e^- \rightarrow Y ( 4 S ) \rightarrow B^+ B^- , B^0 \bar{B}^0$.}
\label{e+e-}
\end{figure}

However, the experimental situation concerning flavour physics
is drastically changing.  
There are several $B$ physics experiments successfully
running at  the moment and,  in the upcoming years, 
new facilities will start
to explore  $B$ physics with increasing 
sensitivity and within various  different  
experimental settings:
Apart from the CLEO experiment (Cornell, USA),
located at the Cornell Electron-Positron Storage Ring (CESR) \cite{cleoexpo}, 
two $B$ factories, 
operating at the  $\Upsilon(4S)$ resonance in an asymmetric mode 
(fig.\ref{e+e-}), have started successfully: 
the BaBar experiment at SLAC 
(Stanford, USA) 
\cite{babarexpo}  
and the BELLE experiment 
at KEK (Tsukuba, Japan) 
\cite{belleexpo}.
Besides the successfully running  hadronic 
$B$ physics program at FERMILAB (Batavia, USA) 
\cite{fermilabexpo} 
there are independent $B$ physics experiments  planned at the
hadronic colliders: the LHC-$B$ experiment at
CERN in Geneva 
\cite{lhcbexpo} 
and the $B$TeV experiment at FERMILAB
\cite{btevexpo}. 
The main motivation for a $B$ physics program at hadron colliders 
is the huge $b$ quark production cross section with  respect to  the one at 
$e^+ e^-$ machines.

While the time of the electroweak precision physics focusing on the 
{\it gauge} sector of the SM  draws to 
a close with the completion of the 
LEP experiments at CERN and the SLC experiment  in Stanford, the era of 
precision flavour physics focusing on the {\it scalar} sector of the SM 
has just begun  with the start of the $B$ factories.

The $B$ system represents an ideal framework for the study of
flavour physics. Since the $b$ quark mass is much larger than the typical
scale of the strong interaction, long-distance strong interactions are
 generally  
less important and are under better control than in kaon physics
thanks to the heavy mass expansion.
Thus, for example  the CP violation  in the $B$ system 
will yield an important independent test of the SM description of 
CP violation (see \cite{Durham}).
$B$ meson decays also allow for a rich CKM phenomenology  
and a stringent test of the unitarity constraints. 

The so-called rare decays  are of particular interest.
These processes represent flavour changing neutral currents (FCNC)
 and occur in the SM  only at the quantum level.
The inclusive rare decay modes are theoretically  clean 
observables because no specific model is needed to describe the 
 hadronic final states. Their role is twofold:  
on the one hand they are relevant to the determination of CKM matrix
elements. On the other hand they are particularly sensitive
to new physics beyond the SM, since additional contributions to the 
decay rate, in which SM particles are replaced by new particles such as the
supersymmetric  charginos or gluinos, are not suppressed by additional
factors $\alpha/(4\pi)$ relative to the SM contribution. This makes 
it possible to observe new physics indirectly -  a strategy 
\mbox{complementary} to the direct production of new (supersymmetric) 
particles. The latter production is reserved for the planned 
hadronic machines such as the LHC at CERN, while the indirect search of the 
$B$ factories  already implies significant restrictions for the parameter 
space of supersymmetric models and will thus lead to important 
clues for the direct search of supersymmetric particles.
It is even possible that these rare processes lead to the first 
evidence of new physics by a significant deviation from the SM
prediction, for example in the observables concerning direct 
 CP  violation, although it will then be difficult to identify
in this way the new structures in detail. But also in the long run,
after new physics has already been discovered, these decays will
play an important role in analyzing in greater detail the 
underlying new dynamics.

Although the general focus within flavour physics is 
at present on $B$ systems, kaon physics 
offers interesting complementary opportunities in the new 
physics search such as the rare
decays $K^+ \rightarrow \pi^+ \nu \bar{\nu}$ and 
 $K_L \rightarrow \pi^0 \nu \bar{\nu}$. 
They are  specifically interesting in view of the current experiments 
at the Brookhaven 
laboratory 
(USA) and suggested experiments at FERMILAB (USA) and at KEK (Japan).

The paper is organized as follows: in Section 2 we briefly discuss 
the role of the strong interaction within flavour physics.
In Section 3 the status of rare $B$  
decays within the SM is reviewed.
In Section 4 we explore the implications of these decays 
for our search of physics beyond the SM. 
In Section 5 we 
discuss direct CP violation  
and in  Section 6 the 
complementary role of rare kaon decays within precision 
flavour physics. In Section 7 we present our summary.

\begin{figure}
\begin{center}
\epsfig{figure=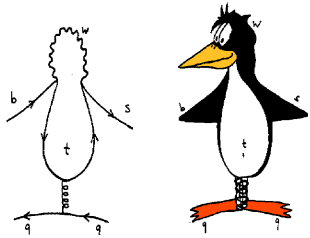,width=4cm}
\caption{Penguin decays of $B$ mesons.}
\label{penguin}
\end{center}
\end{figure}

\setcounter{equation}{0}
\section{Strong interaction in $B$ decays} 
\label{strong}

Flavour physics is governed by the interplay of strong and weak interactions. 
One of the main difficulties in examining the observables in 
flavour physics is the influence of the strong interaction.
As is well known, for matrix elements dominated by long-distance strong 
interactions there is no
adequate quantitative solution available in quantum field theory.
The resulting hadronic uncertainties restrict the opportunities in
flavour physics significantly. The present discussion on the new $g-2$ muon
data \cite{muondata}  also reflects this issue. While the hadronic 
self-energy contribution 
to the $g-2$ observable can be determined by experimental data, the well-known
light-by-light contribution can only be modelled at present (see for example 
\cite{Melnikov}).

However, there are several fundamental tools available,  which 
are directly based on QCD.
High hopes for precise QCD predictions are placed on lattice gauge theoretical 
calculations.  
While there are competitive predictions from lattice gauge theory for
form factors of semileptonic decays, pure hadronic decays  
are less accessible to these methods (\cite{Latt}).
With the help of the  so-called QCD sum rules, a  consistency test between 
hadron physics and perturbative QCD, it becomes
possible to connect hadronic and fundamental QCD parameters directly.
Theoretical predictions via QCD sum rules, however, always have 
relatively large uncertainties \cite{SUM}.
Another approach is the method of factorization \cite{Fakt1}. This method
has recently been systematized for nonleptonic decays
in the heavy quark limit
\cite{Fakt2}.  
However, within this approach  a 
quantitative method to estimate the  $1/m_b$ corrections to this limit 
is missing. The latter contributions can be specifically large
if they are chirally enhanced \cite{Fact3}.
Further fundamental methods whose applications and precision are also 
somewhat restricted are  chiral perturbation theory \cite{CHIRAL}
and heavy quark effective theory \cite{HQET}.

In view of this, the goal must be to minimize theoretical uncertainties
with the help of an optimized combination of different
fundamental methods  solely based on  QCD. 
This can only be done for a selected number of observables in flavour
physics. However, there are also observables, dominated by purely
perturbative contributions, which will make precision
flavour physics possible in the near future. 
Among them inclusive rare $B$ decays (see fig. \ref{penguin} \cite{lenz}) 
play  the most important role.

Inclusive decay modes are theoretically clean and represent 
a theoretical laboratory of perturbative QCD.
In particular,  the decay width $\G(B \to X_s \gamma)$ 
is well approximated by the partonic decay rate
$\G(b\to s \gamma)$, which can be
analysed in renormalization group improved 
perturbation theory:
\begin{equation}
\Gamma ( B \rightarrow X_s \gamma) = \Gamma ( b \rightarrow s \gamma ) +
\Delta^{nonpert.}  
\end{equation}
Nonperturbative effects, $\Delta^{nonpert.}$, play a subdominant role
and are under control thanks to the heavy mass expansion.

Thus, in general,  inclusive decay modes 
should be preferred to exclusive ones from 
the theoretical point of view.
The inclusive modes $B \rightarrow X_{s \, (d)} \gamma$
and $B \rightarrow X_{s\, (d)} l^+l^-$ can be measured by the 
electron-positron experiments ($B$ factories, CLEO) with their kinematic 
constraints and their 
low background, while thy are more difficult to measure
 at hadronic machines.
Exclusive decay modes, however, are
more accessible to experiments, in particular at hadronic machines. 
But in contrast 
to the inclusive modes, they have in general large  nonperturbative 
QCD contributions. 
 Exclusive decays such as $B_{d,s} \rightarrow \mu^+ \mu^-$, 
$B_d \rightarrow K^* \gamma$ and \mbox{$B_d \rightarrow K^* \mu^+ \mu^-$} 
are distinguished observables at the LHC-$B$ experiment.

\begin{figure}
\begin{center}
\epsfig{figure=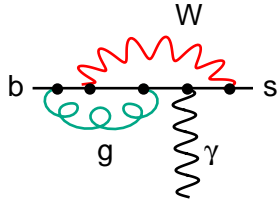,width=4cm}
\end{center}
\vspace{-1cm}
\caption{QCD corrections to the decay $B \rightarrow X_s \gamma$.}
\label{QCDfigure}
\end{figure}

Within inclusive $B$ decay modes, 
short-distance QCD effects lead to a tremendous rate enhancement.
These effects  are induced by 
hard gluon exchange between the quark lines of the one-loop electroweak
diagrams (fig. \ref{QCDfigure}).

The QCD radiative corrections  bring in large logarithms of the form 
$\alpha_s^n(m_b)$ \, $ \log^m(m_b/M)$,
where $M=m_t$ or $M=m_W$ and $m \le n$ (with $n=0,1,2,...$).
This is a natural  feature in any process where two different mass scales
are present.
In order to get a reasonable result at all, one has  to resum at least
the leading-log (LL) series 
\begin{equation}
\alpha_s^n(m_b) \,  \log^n(m_b/M), \,\, (LL)
\label{LLQCD}
\end{equation}
with the help of renormalization
group techniques.
Working to next-to-leading-log (NLL) precision means that one is also 
resumming all the
terms of the form 
\begin{equation}      
\a_s(m_b) \, \left(\a_s^n(m_b) \, \log^n (m_b/M)\right), \,\,\, (NLL).
\label{NLLQCD}
\end{equation}

\begin{figure}
\begin{center}
\epsfig{figure=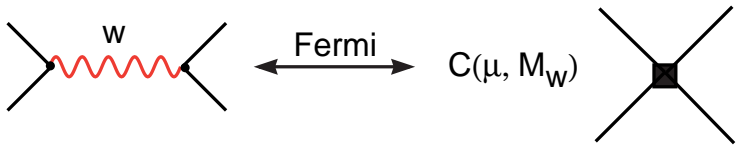,width=10cm}
\end{center}
\vspace{-1.3cm}
\caption{Operator product expansion.}
\label{operatorproductexpansion}
\end{figure}

A suitable framework to achieve the necessary resummations 
of the large logs is an  effective 
low-energy theory with five quarks, obtained by integrating out the
heavy particles, which, in the SM, are the top quark and the $W$ boson. 
The standard method of the operator product expansion allows for a separation
of an amplitude of a weak meson decay process into two distinct parts, 
the long-distance contributions contained 
in the operator matrix elements and 
the short-distance physics described by the so-called Wilson coefficients
 (see fig. \ref{operatorproductexpansion}). 
In the case of $B$ decays, 
the $W$ boson and the top quark with mass 
bigger than the factorization scale are 
integrated out, that is removed from the 
theory as dynamical variables. 
The effective hamiltonian can be written
\begin{equation}
 H_{eff} = - \frac{4 G_{F}}{\sqrt{2}} \, 
\sum  {C_{i}(\mu, M_{heavy})}\,\, \, {\cal O}_i(\mu)
\end{equation}
where ${\cal O}_i(\m)$ are the relevant operators and 
$C_{i}(\mu, M_{heavy})$ are the corresponding Wilson coefficients.
As the heavy fields are integrated out, the complete top and
$W$ mass dependence is contained in the Wilson coefficients.
Working out a convenient set of quantities, both in the effective
(low-energy) theory and in the full (standard model) theory, and 
requiring equality (matching)
up to terms suppressed by higher powers of $m_W$ or $m_t$,
these coefficients can be determined.

Within this framework QCD corrections for the decay rates are 
twofold: the 
ingredients are  the order $\alpha_s$ corrections to the matrix elements
of the various operators and the order $\alpha_s$ 
corrections to the Wilson coefficients, of course both 
at the low-energy scale $\mu \approx  m_b$.
Only the sum of the two 
contributions is renormalization scheme and scale 
independent; in fact, 
from the $\mu$-independence of the effective Hamiltonian,
one can derive a renormalization group equation 
(RGE) for the Wilson 
coefficients $C_i(\mu)$:
\be
\label{RGEa}
\mu \frac{d}{d\mu} C_i(\mu) = \gamma_{ji} \, C_j(\mu) \quad ,
\ee  
where the matrix $\gamma$ is the anomalous dimension
matrix of the operators ${\cal O}_i$. Then there are the 
following three principal steps leading to the
leading-log (next-to-leading-log) result within the
effective field theory approach:
\begin{itemize}
\item {Step 1:\, } One has to match the full SM theory
with the effective theory at the scale $\m=\mu_W$, where
$\mu_W$ denotes a scale of order $m_W$ or $m_t$. At this scale,
the matrix elements of the operators  in the 
effective theory lead to the  same logarithms  as the full theory
calculation. 
Consequently, the Wilson coefficients 
$C_i(\mu_W)$ only pick up small QCD corrections,
which can be calculated in fixed-order perturbation theory.
In the LL (NLL)  program, the matching has to be worked out at the 
$O(\a_s^0)$  ($O(\a_s^1)$) level. 
\item {Step 2:\, } Then one performs the   
evolution of these Wilson coefficients from 
$\m=\mu_W$ down to $\m = \mu_W$, where $\mu_b$ is of the order of $m_b$.
As the matrix elements of the operators evaluated at the low scale
$\mub$ are free of large logarithms, the latter are contained in resummed
form in the Wilson coefficients. For a LL (NLL) calculation, this RGE step
has to be done  using the anomalous dimension matrix up 
to order $\a_s^1$  ($\a_s^2$).
\item {Step 3:\, } To LL (NLL) precision, the 
corrections to the matrix elements 
of the operators $\bra s \g |{\cal O}i (\mu)|b \ket$ at the scale  $\mu = \mub$
have to be calculated to order $\a_s^0$ ($\a_s^1$) precision.
\end{itemize}
Finally, we stress that the step from the leading (LL) to
the next-to-leading (NLL) order within the framework of the 
renormalization group improved perturbation theory is not only a
quantitative one increasing the precision of the theoretical prediction,
 but also a qualitative one, which tests the validity of the perturbative
approach in the given problem.

\setcounter{equation}{0}
\section{Inclusive decay modes}
\label{inclusivesection}

\subsection{Experimental data on $B \rightarrow X_s \gamma$}
\label{experimental}

Among inclusive rare $B$ decays, the $B \rightarrow X_s \gamma$ mode
is the most prominent 
because it is the only one that is
already measured:
in 1993, the first evidence for a penguin-induced $B$ meson decay was found
by the CLEO collaboration. At CESR,
they measured the exclusive 
electromagnetic penguin process $B \to K^* \gamma$.
The inclusive analogue $B \to X_s \gamma$ was also found by the CLEO 
collaboration through the measurement of its characteristic photon 
energy spectrum in 1994 (see \cite{experiment1}).
As this process is dominated by the two-body 
decay $b \to s \gamma$, its photon energy spectrum 
is expected to be a smeared delta function centred 
at $E_\gamma \approx m_b/2$, 
where the smearing is due to perturbative gluon bremsstrahlung and  to 
the nonperturbative Fermi motion of the $b$ quark within the $B$ meson.
Only the high part of the photon energy spectrum
is sensitive to the rare decay $B \to X_s \gamma$.
Some lower cutoff in the photon energy has to be imposed in order to exclude 
the background, mainly  from the nonleptonic charged 
current processes  $b \to c q \bar{q}' + \gamma$ or
$b \to u q \bar{q}' + \gamma$, which have a typical bremsstrahlung 
spectrum that is maximal at $E_\gamma =0$ and falls off for 
larger values of $E_\gamma$. 
Therefore only the ``kinematic'' branching ratio for
$B \to X_s \gamma$ in the range between 
$E_\gamma=2.2$ GeV and the kinematic endpoint at $E_\gamma= 2.7$
GeV could be measured directly.
To obtain from this measurement the total  branching ratio, one has to know
the fraction $R$ of the $B \to X_s \gamma$ events with
$E_\gamma \ge 2.2$ GeV.
This was done in \cite{AG91} where 
the Fermi motion of the $b$ quark in the $B$ meson was
taken into account by using the phenomenological model of Altarelli et
al. (ACCMM model) \cite{ACCMM}.
Using this {\it theoretical} input regarding the photon energy spectrum, 
the value  $R=0.87 \pm 0.06$ was used  
by the CLEO collaboration, leading to the CLEO branching ratio
\cite{CLEOincl}
\begin{equation} 
\label{cleoincl}
{\cal B}(B \to X_s \gamma) = (2.32 \pm 0.57_{stat} \pm 0.35_{sys}) \times 10^{-4}. 
\end{equation}

The first error is statistical and the second 
is systematic (including model dependence). 
This measurement was based on a sample of $2.2 \times 10^6  B\bar B$
events. 

In 1999, CLEO has presented an improved measurement
\cite{CLEOneu},
which is based on $53\%$ more data ($3.3 \times 10^6$ events).
 They also used the
slightly wider $E_{\gamma}$ window starting at $2.1$ GeV. 
The relative error drops almost by a factor of $\sqrt{3}$:
\beq 
\label{cleoneu}
{\cal B}(B \to X_s \gamma) = (3.15 \pm 0.35_{stat} \pm 0.32_{sys} \pm 0.26_{mod}) 
\times 10^{-4}. 
\eeq
The errors represent statistics, systematics, and
the model dependence, respectively.

There are also data at the $Z^0$ peak from the LEP experiments.
The ALEPH collaboration \cite{ALEPH} has measured the
inclusive
branching ratio
\beq
\label{braleph} 
{\cal B}(H_b \to X_s \gamma) = (3.11 \pm 0.80_{stat} \pm 0.72_{sys}) \times 10^{-4}. 
\eeq
It should be noted that
the branching ratio in (\ref{braleph}) involves a  weighted
average of the $B$ mesons and $\Lambda_b$ baryons produced in $Z^0$ decays
(hence the
symbol $H_b$) different from the corresponding one given by CLEO, 
which has been
measured at  the $\Upsilon(4S)$ resonance.
High luminosity is more difficult to obtain
at higher $e^+e^-$ collision energies. Thus, $B\bar{B}$ samples
obtained by the LEP experiments are smaller than the one 
at CESR.  The rate measured by ALEPH
is consistent with the CLEO measurement.

Recently, CLEO presented a refined preliminary analysis (with 
an lower photon energy cut $E_{\gamma} \ge 2.0$ GeV) 
\cite{Moriondinclusive}:
\beq
{\cal B}(B \to X_s \gamma)=(2.85 \pm 0.35_{stat} \pm 0.22_{sys}) 
\times 10^{-4} 
\label{Cleomoriond}
\eeq
and also BELLE has presented preliminary data \cite{Moriondinclusive}
 of competitive experimental accuracy: 
\beq
\label{Bellemoriond}
{\cal B}(B \to X_s \gamma) = (3.37 \pm 0.53_{stat} \pm 0.42_{sys} 
\pm 0.54_{mod}) \times 10^{-4}.
\eeq
More accurate data can be expected in the near future.
With the expected high luminosity of the $B$-factories, 
an experimental accuracy below $10\%$ 
in the inclusive $B \rightarrow X_s \gamma$ mode appears 
to be possible.

The uncertainty regarding the fraction $R$ of the $B \to X_s \gamma$ 
events with
$E_\gamma \ge 2.2$ GeV quoted in the experimental measurement, also
cited as model dependence, should 
be regarded as a purely theoretical uncertainty.
As mentioned above, 
the fraction $R$ was calculated 
in \cite{AG91} using the phenomenological model by Altarelli et
al., where the Fermi motion of the $b$ quark in the $B$ meson 
is characterized by two parameters, the
average Fermi momentum $p_F$ of the $b$ quark and the mass $m_q$ of the 
spectator quark.
The error on the fraction $R$ is essentially obtained by varying 
the model parameters $p_F$ and $m_q$ in the range for which the ACCMM
model correctly describes the energy spectrum of the charged lepton in
the semileptonic decays $B \to X_c \ell \nu$ and $B \to X_u \ell \nu$,
measured by CLEO and ARGUS. 
In \cite{AG91} a first comparison between the calculated photon energy 
spectrum 
and the one measured by the CLEO collaboration was presented.
The (normalized) measured
photon energy spectrum and the theoretical one are in agreement for those 
values of  $p_F$ and
$m_q$, that correctly describe the inclusive semileptonic CLEO data $B \to X_c
\ell \nu$ and $B \to X_u \ell \nu$;
at present, the data from the
radiative decays is, however, not precise enough to 
further constrain the values of $p_F$ and $m_q$. 
The best fit between 
the theoretical and measured photon energy
spectrum is obtained for $p_F=450$ MeV and $m_q=0$.
One should mention that   the analysis \cite{AG91} of the
photon energy spectrum, in particular 
the calculation of the fraction $R$ in the 
ACCMM model used by CLEO, does not include the full
NLL information, which becomes available in the meantime.

\begin{figure} 
\begin{center}
\epsfig{figure=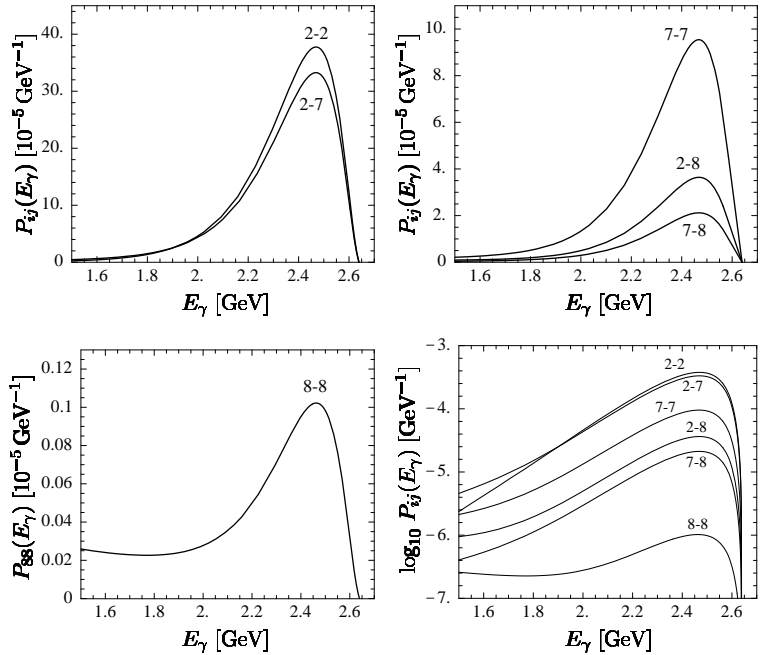,width=10cm}
\end{center}
\label{photonspectrum}
\vspace{-0.5cm}
\caption{Different components of the photon spectrum in the $B\to X_s\gamma$
decay, from \cite{Kagan}.}
\label{photonspectrumkagan}
\end{figure}

Besides this phenomenological model by Altarelli et al.,
 more fundamental theoretical methods 
are  available today to implement the bound state effects, 
namely by making use 
of operator product expansion techniques in the framework of heavy quark 
effective theory (HQET).
A new analysis along these lines  was presented in \cite{Kagan}.
Unfortunately, the operator product expansion breaks down 
near the endpoint of the photon energy  spectrum; therefore, an infinite 
number of leading-twist corrections  has to be resummed into
a nonperturbative universal shape function, which determines the
light-cone momentum distribution of the $b$-quark in the $B$ meson
 \cite{shapefunction}. The 
physical 
decay distributions are then obtained from a convolution of 
parton model spectra 
with this shape function.
At present this function cannot be calculated, but there is at least some 
information 
on the moments of the shape function, which are related to the forward matrix 
elements 
of local operators. Ans\"atze  for the shape 
function, constrained by the latter information, are used. 
In contrast to the older analysis based on the ACCMM model, 
the new analysis of Kagan and Neubert \cite{Kagan}
includes the full NLL information. Their  
fraction
$R=0.78^{+0.09}_{-0.11}$ (for the energy cut $E_\gamma > 2.2$ GeV) 
is 
smaller than the factor used by CLEO.

An important observation is that the shape of the photon spectrum 
is not sensitive to physics beyond the SM. 
As can be seen in fig. \ref{photonspectrumkagan},
all different contributions to the spectrum 
(corresponding to the interference
terms of the various operators involved)
have a very similar
shape besides the small $8$-$8$ contribution. 
This implies that we do not have to assume the
correctness of the SM in the experimental analysis and, 
thus, a precise measurement of the photon spectrum  can be used 
to determine the parameters of the shape function.

Clearly, a lower experimental cut decreases the sensitivity 
to the parameters of the shape function (or, more generally, the 
model dependence). With respect to this, the ideal energy cut
would  be $1.6$ GeV. However, in this case a better 
understanding of the  $\psi$ background would be mandatory.
The intermediate $\psi$ background, namely 
${B} \to \psi X_s$ followed by $\psi \to X' \gamma$ is  
more than $4 \times 10^{-4}$ in the `total' branching ratio.  
With the present energy cut of $2.1$ GeV this contribution is
suppressed and estimated to be less than $5 \%$ \cite{Misiakichep}.

Another future aim should be to determine the shape function
(and analogously the parameter of the ACCMM model) 
by using the high-precision measurements of the photon  energy
spectrum.

\subsection{NLL QCD calculations}
\label{NLLQCDcorrections}

As mentioned above, the inclusive decay $B \rightarrow X_s \gamma$ is a
laboratory for perturbative QCD. 
Nonperturbative effects (see section \ref{sectionnonpert}) 
play a subdominant role and are well  under control thanks to 
the heavy quark expansion. 
The short-distance  QCD corrections enhance the 
partonic decay rate $ \Gamma(b \to s \g)$  by more than a factor of $2$.
The corresponding large  logarithms of the form 
$\alpha_s^n(m_b) \, \log^m(m_b/M)$,
where $M=m_t$ or $M=m_W$ and $m \le n$ (with $n=0,1,2,...$), have to
be summed with the help of the renormalization group improved perturbation 
theory as presented in section \ref{strong}.

\begin{figure}
\begin{center}
\epsfig{figure=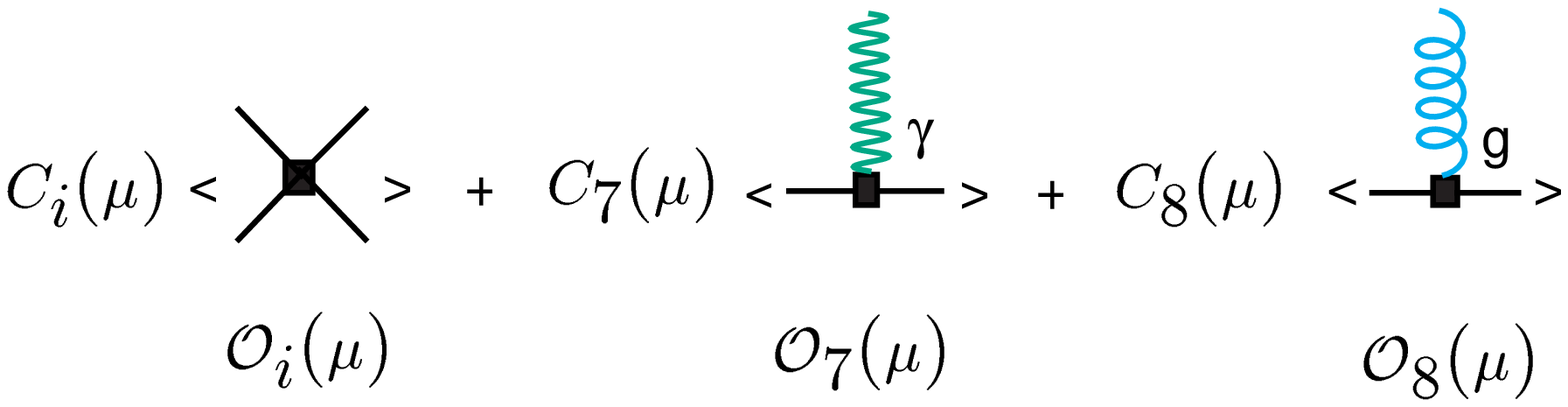,width=10cm}
\end{center}
\vspace{-0.5cm}
\caption{SM Hamiltonian in the case of $B \rightarrow X_s \gamma$.}
\label{SMhamiltonian}
\end{figure}

The effective Hamiltonian relevant to  $B \to X_s \gamma$ 
in the SM reads 

\begin{equation}
\label{heff}
H_{eff}(B \to X_s \gamma)
       = - \frac{4 G_{F}}{\sqrt{2}} \, \lambda_{t} \, \sum_{i=1}^{8}
C_{i}(\mu) \, {\cal O}_i(\mu) \quad, 
\end{equation}
where ${\cal O}_i(\m)$ are the relevant operators, 
$C_{i}(\mu)$ are the corresponding Wilson coefficients,
which contain the complete top- and $W$- mass dependence
(see fig. \ref{SMhamiltonian}),
and $\lambda_t=V_{tb}V_{ts}^*$ with $V_{ij}$,  the
CKM matrix elements. The CKM dependence globally factorizes,
because we work in the approximation $\lambda_u=0$ 
(in the case of $B \to X_s \gamma$).
One neglects the operators with dimension $>6$. which are suppressed 
by higher powers of $1/m_{W}$. 

Using the equations of motion for the operators, one arrives at the 
following basis  of dimension-6 operators: 
\bea
\label{operators}
{\cal O}_1 &=& \left( \bar{c}_{L \beta} \g^\m b_{L \alpha} \right) \,
        \left( \bar{s}_{L \alpha} \g_\m c_{L \beta} \right)\,, \nonumber \\
{\cal O}_2 &=& \left( \bar{c}_{L \alpha} \g^\m b_{L \alpha} \right) \,
        \left( \bar{s}_{L \beta} \g_\m c_{L \beta} \right) \,,\nonumber \\
{\cal O}_3 &=& \left( \bar{s}_{L \alpha} \g^\m b_{L \alpha} \right) \, \left[
        \left( \bar{u}_{L \beta} \g_\m u_{L \beta} \right) + ... +
        \left( \bar{b}_{L \beta} \g_\m b_{L \beta} \right) \right] \,,
        \nonumber \\
{\cal O}_4 &=& \left( \bar{s}_{L \alpha} \g^\m b_{L \beta} \right) \, \left[
        \left( \bar{u}_{L \beta} \g_\m u_{L \alpha} \right) + ... +
        \left( \bar{b}_{L \beta} \g_\m b_{L \alpha} \right) \right] \,,
        \nonumber \\
{\cal O}_5 &=& \left( \bar{s}_{L \alpha} \g^\m b_{L \alpha} \right) \, \left[
        \left( \bar{u}_{R \beta} \g_\m u_{R \beta} \right) + ... +
        \left( \bar{b}_{R \beta} \g_\m b_{R \beta} \right) \right] \,,
        \nonumber \\
{\cal O}_6 &=& \left( \bar{s}_{L \alpha} \g^\m b_{L \beta} \right) \, \left[
        \left( \bar{u}_{R \beta} \g_\m u_{R \alpha} \right) + ... +
        \left( \bar{b}_{R \beta} \g_\m b_{R \alpha} \right) \right] \,,
        \nonumber \\
{\cal O}_7 &=& (e/16\p^{2}) \, \bar{s}_{\alpha} \, \sigma^{\m \n}
      \, ({\overline m}_{b}(\mu)  P_R + {\overline m}_s(\mu)  P_L) \, b_{\alpha} \ F_{\m \n} \,,
        \nonumber \\
{\cal O}_8 &=& (g_s/16\p^{2}) \, \bar{s}_{\alpha} \, \sigma^{\m \n}
      \, ({\overline m}_{b}(\mu)  P_R + {\overline m}_{s}(\mu)  P_L) \, 
T_{\alpha\beta}^A  \,b_{\beta}
      \ G^A_{\m \n}.
\eea
In the dipole type operators ${\cal O}_7$ and ${\cal O}_8$, 
$e$ and $F_{\m \n}$ ($g_s$ and $G^A_{\m \n}$)
denote the electromagnetic (strong)
coupling constant and field strength tensor,  respectively; 
$T^a$ ($a=1,8$) denote $SU(3)$ colour generators.

The error of the leading logarithmic (LL) result \cite{counterterm}
was  dominated by a large renormalization scale dependence 
at the $\pm 25\%$ level, which alreday indicated
the importance of the NLL series.
By convention, the dependence on the renormalization scale $\mu_b$ is
obtained by the variation $m_b/2 < \mu_b < 2 m_b$.
The former measurement of the CLEO collaboration (see (\ref{cleoincl})) 
overlaps with the estimates based on LL  calculations,
and the experimental and 
theoretical errors are comparable.
In view of the expected increase in the experimental precision
in the near future, it became clear that a systematic inclusion of 
the NLL corrections
was becoming necessary. Moreover, such a NLL program 
is also important in order to ensure the validity
of renormalization group improved perturbation theory in this specific 
phenomenolgical application.

This ambitious NLL enterprise was completed some years ago.
This was a joint effort of many different groups 
(\cite{AG91},\cite{Pott}, \cite{GHW}, 
\cite{Adel}, \cite{Mikolaj}),\cite{GH}).
The theoretical error of the  previous 
LL result was substantially reduced 
to $\pm 10\%$, and the central value of the partonic
decay rate increased by about $20\%$.

All three steps to NLL precision listed below (\ref{RGEa})
involve rather difficult calculations.   

\begin{itemize}

\item The most difficult part in Step 1 is the 
two-loop (or  order $\a_s$) matching of the dipole operators ${\cal O}_7$ and 
${\cal O}_8$. 
It involves two-loop diagrams both in the full and in the effective theory. 
It was first worked out by Adel and Yao \cite{Adel}. 
As this is a crucial step in the NLL program,
Greub and Hurth confirmed their findings in a detailed re-calculation 
using a different method \cite{GH}.
Two further recalculations of this result 
\cite{CDGG,INFRARED} were presented in the meanwhile, confirming
the original results in \cite{Adel}. 
In order to match the dimension-6 operators ${\cal O}_7$ and ${\cal O}_8$, 
it is sufficient to
extract the terms  of order $m_b \, \frac{m_b^2}{M^2} $ ($M=m_W,m_t$)
from the SM 
matrix elements for $b \to s \g$ and $b \to s g$.
Terms suppressed by additional powers of
$m_b/M$ correspond to higher-dimensional operators in the effective theory.
In \cite{GH} the finite parts of the 
two-loop diagrams in the SM were calculated by means of the 
well-known method of aymptotic mass expansions, which naturally leads 
to a systematic expansion of Feynman diagrams in inverse powers of
$M$.

\item 
The order $\a_s^2$ anomalous dimension matrix (Step 2) has been 
worked out
by Chetyrkin, Misiak and M\"unz \cite{Mikolaj}. 
In particular, the calculation of the elements $\gamma_{i7}$ 
and $\gamma_{i8}$ ($i=1,...,6$) in the $O(\a_s^2)$ anomalous
dimension matrix involves a huge number of three-loop diagrams
from which the pole parts (in the $d-4$ expansion) have to be extracted.
This extraction was
 simplified by a clever decomposition of the scalar propagator.
Moreover, the number of necessary evanescent operators was reduced by 
a new choice of a basis of dimension-6 operators.   
Using the matching result (Step 1), these authors obtained 
the NLL  correction to the Wilson coefficient $C_7(\m_b)$.
Numerically, the LL and the NLL value 
for $C_{7}(\m_b)$ 
turn out to be  rather similar; the NLL 
corrections to the Wilson coefficient $C_7(\m_b)$ 
lead to a change of the  $B \to X_s \gamma$
decay rate that does not exceed  6\% \cite{Mikolaj}.

It should be stressed that the result of Step 2, in particular  
the entries $\gamma_{i7}$  and $\gamma_{i8}$ ($i=1,...,6$)
of the anomalous dimension matrix to NLL precision, is the only 
part of the complete  NLL enterprise which has not been confirmed 
by an independent group.

\item
Step 3 basically consists of bremsstrahlung corrections and virtual
corrections. While the bremsstrahlung corrections
were worked out some time ago by Ali and Greub \cite{AG91} and were
 confirmed and extended by Pott \cite{Pott}, a
complete analysis of the virtual two-loop corrections (up to the contributions 
of the four-quark operators with very small coefficients) was presented
by Greub, Hurth and Wyler \cite{GHW}. 
The latter  calculation involves two-loop diagrams, where the full charm 
dependence has to be taken into account.   
By using Mellin-Barnes techniques in the Feynman parameter integrals, 
the result of these two-loop
diagrams was obtained in the form
\be
\label{Mellin}
c_0 + \sum_{n=0,1,2,...;m=0,1,2,3} c_{nm} \left( \frac{m_c^2}{m_b^2}
\right)^n \, \log^m \frac{m_c^2}{m_b^2} \, ,
\ee
where the quantities $c_0$ and $c_{nm}$ are independent of $m_c$.
The convergence of the Mellin-Barnes series was proved;
the practical convergence of the series (\ref{Mellin}) was
also checked explicitly. 
Moreover, a finite result is obtained in the limit $m_c \to 0$,
as there is no naked logarithm of $m_c^2/m_b^2$. This observation
is of some
importance in the $b \to d \gamma$ process, where the $u$-quark
propagation
in the loop is not CKM-suppressed (see below).
The main result of Step 3  consists in a drastic reduction of the 
renormalization 
scale uncertainty from about $\pm 25\%$ to about $\pm 6\%$.
The central value was shifted by about $20 \%$. 

In \cite{GHW} these results are presented also in the 
't Hooft-Veltman scheme, which may be regarded
as a first step towards a cross-check 
of the complete  NLL calculation  prediction in a different 
renormalization  scheme.

Quite recently, the results of the matrix elements based
on the operator ${\cal O}_2$ 
were confirmed by an
independent group \cite{Burasnew} with the help of the method
of  asymptotic expansions.

\end{itemize}

\begin{figure} 
\centerline{
\psfig{figure=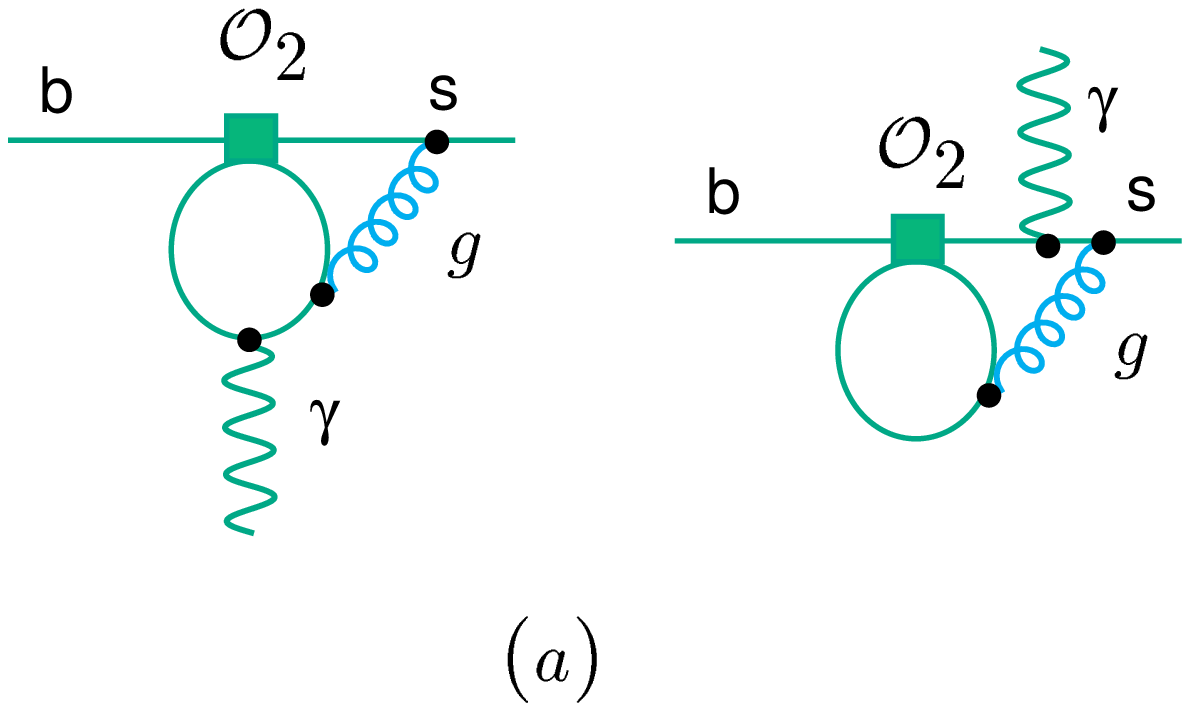,height=4cm} 
\hspace{1cm}\psfig{figure=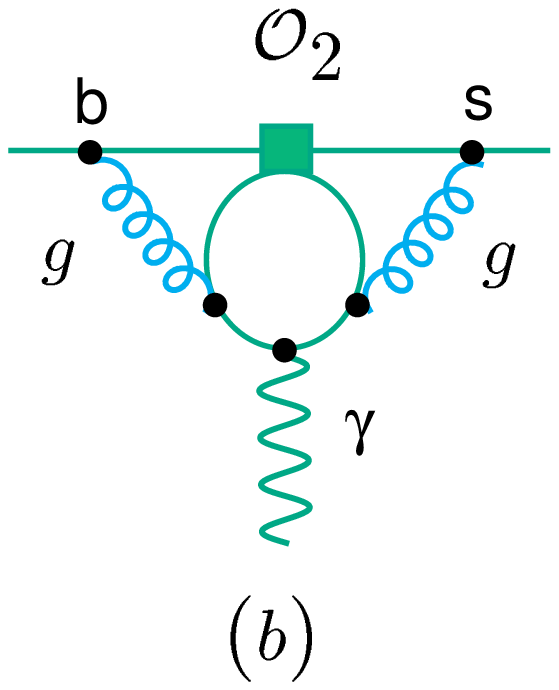,height=4cm}} 
\caption[]{(a) Typical diagrams (finite parts) contributing 
to the matrix element of the operator ${\cal O}_2$ at the NLL level (Step3);
(b) typical diagram (infinite part) contributing to the NLL anomalous 
dimension matrix (Step2); typical diagram (finite part) contributing 
in the NLL matching calculation shown in fig.\ref{QCDfigure} (Step1).
\label{feynman}}
\end{figure}

It is clear that many parts of the NLL calculations at the
partonic level in the  case of $b \rightarrow s \gamma$ 
can be 
straightforwardly taken over to the cases $b \rightarrow d \gamma$,
$b \rightarrow s \, gluon$ and $b \rightarrow s l^+ l^-$.
In the latter case, however, many modifications are necessary; 
 in particular the operator basis gets enlarged 
as will be discussed below.

Combining the NLL calculations of the three steps, 
the first complete theoretical prediction to NLL  precision 
for the branching ratio of $B \rightarrow X_s \gamma$
was presented in \cite{Mikolaj} (see also \cite{Towards}):
\be
\label{mikolajend}
{\cal B}(B \to X_s \g)=(3.28 \pm 0.33) \times 10^{-4}.
\ee
The theoretical error has two dominant sources.  The 
$\mu$ dependence, which is now reduced to about $6 \%$.
The other main  uncertainty of $5\%$ stems from the $m_c/m_b$ dependence. 
This first theoretical NLL prediction already included the 
nonperturbative correction scaling with $1/m_b^2$, which are rather small
(at the $1\%$ level) (see section \ref{sectionnonpert}).
Surprisingly, these first NLL predictions (\cite{Mikolaj},\cite{Towards}) 
are almost identical to the current prediction quoted in  
(\ref{currentprediction}), 
in spite of so many important additional refinements such as the 
electroweak two-loop corrections and the nonperturbative
corrections which will be discussed below.

\subsection{Nonperturbative contributions} 
\label{sectionnonpert}

Within the framework of the heavy mass expansion,
nonperturbative corrections 
to the branching ratio of decay $B \rightarrow X_s \gamma$ 
can be  singled out. These contributions also apply to the case
of the decay $B \rightarrow X_d \gamma$ and, with some modifications, 
to the case of the decay  $B \rightarrow X_s l^+ l^-$.

If one neglects perturbative QCD corrections and assumes that the decay  
$B \to X_s \gamma$ is due to the operator ${\cal O}_7$ only,
the calculation of the differential decay rate basically amounts
to working  out the imaginary part of the forward scattering
amplitude $T(q)$:
\be
\label{forward}
T(q) = i \, \int d^4x \, \bra B|T {\cal O}_7^+(x) \, {\cal O}_7(0) |B \ket \,
\exp (iqx) \quad .
\ee
Using the operator product expansion for $T {\cal O}_7^+(x) \, {\cal O}_7(0)$
and heavy quark effective theory methods, the decay width
$\Gamma(B \to X_s \gamma)$ reads \cite{Falk,Alineu} 
(modulo higher terms in the
$1/m_b$ expansion):
\bea
\label{width}
\Gamma_{B \to X_s \gamma}^{({\cal O}_7,{\cal O}_7)} &=&
\frac{\a G_F^2 m_b^5}{32 \pi^4} \, |V_{tb} V_{ts}|^2 \, C_7^2(m_b) \,
\left( 1 + \frac{\delta^{NP}_{rad}}{m_b^2} \right) \quad , \nonumber \\
\delta^{NP}_{rad} &=& \frac{1}{2} \lambda_1 - \frac{9}{2} \lambda_2 \quad ,
\eea
where $\lambda_1$ and $\lambda_2$ are the parameters 
for kinetic energy and the chromomagnetic
energy. Using $\lambda_1=-0.5 \, \mbox{GeV}^2$ and 
$\lambda_2=0.12 \, 
\mbox{GeV}^2$, one gets $\delta_{rad}^{NP} \simeq -4\%$.
The $B \rightarrow X_s \gamma$ decay width    is usually 
normalized by the semileptonic one.
The semileptonic decay width gets $1/m_b^2$ corrections,
which are negative; thus, 
the nonperturbative corrections scaling with $1/m^2_b$ tend to cancel
in the branching ratio ${\cal B}(B \to X_s \gamma)$, and only about $1\%$
remains.

\begin{figure} 
\centerline{
\psfig{figure=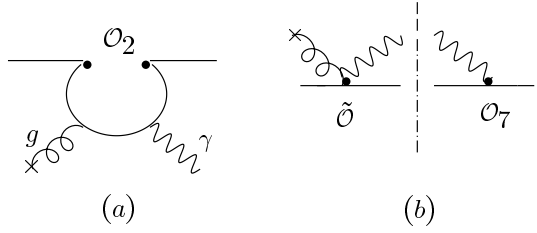,height=3cm}} 
\caption[]{a)Feynman diagram from which the operator $\tilde{\cal O}$
arises. b) Relevant cut-diagram for the $({\cal O}_2,{\cal O}_7)$-interference.
\label{Voloshinfig}}
\end{figure}

Voloshin \cite{Voloshin} considered the nonperturbative
effects when including also the operator ${\cal O}_2$. 
This effect is generated by the diagram in Fig. \ref{Voloshinfig}a 
(and by the one, not shown, where the gluon and the photon
are interchanged); $g$ is a soft gluon interacting with the
charm quarks in the loop. Up to a characteristic Lorentz structure,
this loop is given by the integral
\be
\label{volloop} 
\int_0^1 dx \, \int_0^{1-x} dy \, \frac{xy}{m_c^2-k_g^2 x(1-x) -2xy k_g k_\g}
\quad .
\ee
As the gluon is soft, i.e. $k_g^2,k_g k_\g \approx \Lambda^{QCD} \, m_b/2
\ll m_c^2$, the integral can be expanded in $k_g$. The (formally)
leading operator, denoted by $\tilde{\cal O}$, is
\be
\label{tildeo}
\tilde{\cal O} = \frac{G_F}{\sqrt{2}} V_{cb} V_{cs}^* C_2 \,
\frac{e Q_c}{48 \pi^2 m_c^2} \, \bar{s} \g_\mu (1-\g_5) g_s 
G_{\nu \lambda} b \, \epsilon^{\mu \nu \rho \sigma} \partial^\lambda
F_{\rho \sigma} \quad .
\ee 
Then working out the cut diagram shown in Fig. \ref{Voloshinfig}b,
one obtains the nonperturbative 
contribution $\Gamma^{(\tilde{\cal O},{\cal O}_7)}_{B \to X_s \gamma}$
to the decay width,
which is due to the $({\cal O}_2,{\cal O}_7)$ interference.
Normalizing this contribution by the LL partonic width, one obtains
\be
\label{voleffect}
\frac{\Gamma^{(\tilde{\cal O},{\cal O}_7)}_{B \to X_s \gamma
}}{\Gamma_{b \to s \g}^{LL}} = -\frac{1}{9} \, \frac{C_2}{C_7} 
\frac{\l_2}{m_c^2} \simeq +0.03 \quad .
\ee

As the expansion parameter is $m_b \Lambda_{QCD}/m_c^2 \approx 0.6$
(rather than $\Lambda^2_{QCD}/m_c^2$), it is not a priori clear
whether formally higher order terms in the $m_c$ expansion are
numerically suppressed. More detailed investigations 
\cite{Wise,Peccei,Rey}
have shown that higher order terms are indeed suppressed, because
the corresponding expansion coefficients are small.

The analogous $1/m_c^2$ effect 
has been found independently in the exclusive mode
$B \to K^* \gamma$ in ref. \cite{Wyler}. Numerically, the effect
there is also at the few per cent level.
Moreover, the analysis of the $1/m_c^2$ effects was extended to
the decay $B \rightarrow X_s l^+l^-$ in \cite{Rey,buchallanewnew}. 

As was recently emphasized by Misiak \cite{Misiak:1999jh}, 
an analogous  systematic analysis  of terms like 
$\Gamma^{{({\cal O}_2,{\cal O}_2)}}_{B \to X_s \gamma}$ at  
first order in $\alpha_s$ is still missing. 
Rigorous  techniques such as  operator product expansions 
do not seem to be applicable in this case.

\subsection{Theoretical status}

\subsubsection{$B \rightarrow X_s \gamma$}

The theoretical prediction for the partonic $b \rightarrow s \gamma$ decay 
rate  is usually 
normalized by the semileptonic decay rate in order to get rid of uncertainties 
related with the CKM matrix elements and the fifth power of the $b$ quark mass.
Moreover,  an explicit lower cut on the photon energy in the
bremstrahlung correction is often made: 
\be
R_{quark}(\delta) = 
\frac{\Gamma[ b \to s \gamma]+\Gamma[ b \to s \gamma  gluon]_\delta}{\Gamma[ b \to X_c e \bar{\nu}_e ]}
\ee
where the subscript $\delta$ means that only photons with energy 
$E_{\gamma}>(1-\delta) E_{\gamma}^{max} = (1-\delta) \frac{m_b}{2}$ are counted. 
The ratio $R_{quark}$ is divergent in the limit $\delta \rightarrow 1$ due to the 
unphysical soft photon divergence in the subprocess $b \rightarrow s \gamma gluon$. 
In this limit only the sum of $\Gamma[b \to s \gamma]$,
$\Gamma[b \to s  gluon]$ and $\Gamma[b \to s \gamma gluon]$ is a reasonable 
physical quantity, in which all divergences cancel out. 
In \cite{Kagan} it was shown that 
the theoretical result is
rather sensitive to the unphysical soft-photon divergence;
 the choice $\delta=0.90$ was suggested as the optimized definition 
of the total decay rate. 
In the analysis presented in \cite{GHW} 
the limit $\delta \rightarrow 1$ is taken and 
the singularities are removed by adding the 
virtual photon corrections to $b \rightarrow s gluon$.

It is suggestive to  give up the concept of a 
total decay rate of $b \rightarrow s \gamma$ 
and compare theory and experiment 
using the same energy cut as CLEO  ($E_{\gamma}>2.1$ GeV). Then also  
the theoretical uncertainty  regarding the photon energy spectrum mentioned 
above would occur naturally in the theoretical prediction.

In the meanwhile detailed studies of the electroweak corrections were
performed.  In   \cite{Marciano} 
           part of the electroweak two-loop contributions, 
namely contributions from  fermion loops in gauge boson
propagators ($\gamma$ and $W$) and from short-distance photonic 
loop corrections, were calculated. Moreover, it was  found that  
the on-shell value of the fine
structure constant  $1/\a_{em}=137$ is more appropriate for real
photon emission than 
the value $1/\a_{em} = (130.3 \pm 2.3)$ used in previous analyses. 
The QED loop calculations in \cite{Marciano} confirmed this expectation.
This change in $\a_{em}$ leads to 
a reduction of $5\%$ in $R_{quark}$.
In \cite{Kagan} the QED analysis made in 
\cite{Marciano} was improved
by resumming the contributions of order 
$\alpha \log(\mu_b/M) (\alpha_s \log(\mu_b/M)^n$
to all orders (while in \cite{Marciano} only the $n=0$ contribution
was included). This resummation decreases the QED corrections.
In \cite{Strumia} a complete analysis of the 
heavy top and the heavy Higgs corrections 
in the limit
$m_W \rightarrow 0$ was made. This analysis was recently refined 
in \cite{Gambino:2000fz}. A 
$2 \%$ reduction of the branching ratio of $B \rightarrow X_s \gamma$ 
due to purely electroweak corrections is found.

Using the measured semileptonic branching ratio ${\cal B}^{sl}_{exp.}$,
the  branching ratio ${\cal B}(B \to X_s \gamma)$ is given by 
\be
{\cal B}(B \to X_s \gamma) = R_{quark}  \times {\cal B}^{sl}_{exp.} (1 + \Delta_{nonpert}),
\ee
where the nonperturbative corrections scaling with $1/m^2_b$ and 
$1/m^2_c$, summed in $\Delta_{nonpert}$, 
have a numerical effect of $+1\%$ \cite{Falk,Alineu} and 
$+3\%$ \cite{Voloshin},  respectively, on the branching ratio only.

For a comparison with the ALEPH measurement (\ref{braleph}) 
the measured semileptonic branching ratio ${\cal B}(H_b \to X_{c,u} \ell 
\nu)$ should be used consistently.
 This leads to a larger theoretical prediction for the LEP experiments.

Including only the resummed QED corrections and the nonperturabtive
corrections discussed in section \ref{sectionnonpert},
using the on-shell value of 
$\alpha_{em}$ and working with the 
convention $\delta \rightarrow 1$ in $R_{quark}$, 
one ends up with  the following theoretical prediction for the
$B \rightarrow  X_s \gamma$ branching ratio \cite{greubhurthQCD}: 
\beq
{\cal B}(B \rightarrow X_s \gamma) = (3.32 \pm 0.14 \pm 0.26)\times 10^{-4},
\label{currentprediction}
\eeq
where the first error represents the uncertainty 
regarding the scale dependences, while the second error is the
uncertainty due to the input parameters. In the second error 
the uncertainty due to the parameter $m_c/m_b$ is dominant. 

Quite recently, quark mass effects within the decay 
$B \rightarrow X_s \gamma$ were further analysed \cite{GambinoMisiak}, 
in particular the definitions of the quark masses $m_c$ and $m_b$ 
in the matrix element
$\bra {\cal O}_2 \ket \, \equiv \, \bar X_s \gamma| (\bar{s}c)_{V-A}(\bar{c}b)_{V-A}|b \ket$.
Since  the charm quark in the matrix element $\bra{{\cal O}}_2 \ket$ are 
dominantly off-shell (see fig. \ref{feynman}a)
the authors argue that the running charm mass should be chosen instead
of the pole mass.                  . 
The latter choice was used  
in all previous analyses \cite{GHW,Mikolaj,CDGG,Kagan,greubhurthQCD}.  
\begin{equation}
 m_c^{\rm pole}/m_b^{\rm pole} \qquad 
{{\Rightarrow}}\, \qquad   m_c^{\overline{\rm MS}}(\mu)/m_b^{\rm pole}, \,  \,\,
\mu \in [m_c,m_b].
\end{equation}

Since the  matrix element starts at NLL order and, thus, 
the renormalization scheme for $m_c$ and $m_b$ is an NNLL issue,
one should regard this choice as an educated guess of the NNLL
corrections. However, this new choice is guided by the 
experience gained from many higher-order calculations
in perturbation theory.  
Numerically, the shift from $m_c^{\rm pole}/m_b^{\rm pole} =  0.29 \pm 0.02$
to $m_c^{\overline{\rm MS}}(\mu)/m_b^{\rm pole} = 0.22 \pm 0.04$
is rather important and leads to a $+ 11 \%$ shift of the central 
value of the $B \rightarrow X_s \gamma$ branching ratio.  
The authors of \cite{GambinoMisiak} quote a weighted 
experimental world average using the preliminary data from 
CLEO and BELLE, (\ref{Cleomoriond}) and (\ref{Bellemoriond}),
and the published ALEPH data (\ref{braleph}):
\begin{equation} 
\label{worldaverage}
{\cal B}(B \to X_s \gamma) = (2.96 \pm 0.35) \times 10^{-4}.
\end{equation}
With their new choice of the charm mass and with $\delta = 0.9$, 
their theoretical prediction 
for the `total' branching ratio is 
\begin{equation} 
\label{totalbr}
{\cal B}(B \to X_s \gamma) = (3.73 \pm 0.30) \times 10^{-4},
\end{equation}
which means that the difference between the theoretical and the experimental 
value is consistent with zero at the level of $1.6 \sigma$ 
(if one assumes that a statistical interpretation 
of this difference is really possible). 
Because the choice of the renormalization scheme for $m_c$ and $m_b$ 
is a NNLL effect, one could  argue for a  larger theoretical uncertainty  
in $m_c^{\overline{\rm MS}}(\mu)/m_b^{\rm pole}$ 
which includes also the value of $m_c^{pole}$.
A  more conservative choice would then be  
$m_c^{\overline{\rm MS}}(\mu)/m_b^{\rm pole} = 0.22 \pm 0.07$ which 
would reduce the significance of the perceived discrepancy.

Instead of making a theoretical prediction for the branching ratio
${\cal B}(B \to X_s \gamma)$,
one can use the experimental data  and 
theory in order to directly determine
the combination $|V_{tb} V_{ts}^*|/|V_{cb}|$ of the CKM matrix elements; 
in turn, one can determine 
$|V_{ts}|$ by making use of the relatively well  known
CKM matrix elements $V_{cb}$ and $V_{tb}$. 
An update of the analysis in \cite{aliokt97} was presented 
in \cite{greubhurthQCD}.  
Using the CLEO data (\ref{cleoneu}),
the ALEPH data (\ref{braleph}), and the theoretical prediction 
(\ref{currentprediction}), one finds \cite{greubhurthQCD}
\begin{eqnarray*}
\frac{|V_{ts}^* V_{tb}|}{|V_{cb}|} &=& 0.95 \pm 0.08_{exp.} \pm 0.05_{th.}
\quad \rm{CLEO} \nonumber \\
\frac{|V_{ts}^* V_{tb}|}{|V_{cb}|} &=& 0.91 \pm 0.15_{exp.} \pm 0.04_{th.}    
\quad \rm{ALEPH.} 
\nonumber
\end{eqnarray*}
The average of the two measurements yields
\beq
 \frac{|V_{ts}^* V_{tb}|}{|V_{cb}|} = 0.93 \pm 0.09  \pm 0.03
=0.93 \pm 0.10
\eeq
where in the very last step the theoretical and experimental errors were
added in quadrature.  
Using
$|V_{tb}|=0.99 \pm 0.15$ from the CDF 
measurement and  $|V_{cb}|=0.0393 \pm 0.0028$ 
extracted from semileptonic $B$ decays, one obtains \cite{greubhurthQCD}
\beq
\label{vts}
|V_{ts}|=0.037 \pm 0.007,
\eeq
where all the errors were added in quadrature. This  is probably
the most direct determination of this CKM matrix element.
With an improved measurement of ${\cal B}(B \to X_s \gamma)$
and $V_{tb}$, one expects to reduce the present error on $|V_{ts}|$
by a factor of 2 or even more.

Finally, some remarks on the decay mode 
$b \rightarrow s \, gluon$ are in order. 
The effective Hamiltonian is the same as in the 
$b \rightarrow s \gamma$ case. By replacing the photon
by the gluon, 
the NLL QCD calculation of $b \rightarrow s \gamma$  
can also be used.
However, in the calculation of 
the matrix element  of the operator ${\cal O}_2$,  further diagrams 
with the nonabelian three-gluon coupling have to be calculated
\cite{Liniger}.
Numerically, one  obtains
${\cal B}(b \rightarrow s \, gluon)=(5.0 \pm 1.0) \times 10^{-3}$, which is 
more than a factor of 2 
larger than the former LL result 
${\cal B}(b \rightarrow s \, gluon)=(2.2 \pm 0.8) \times 10^{-3}$ 
\cite{counterterm}.
The mode $b \rightarrow s \, gluon$
represents one component to the  
inclusive charmless hadronic decays, $B \to X_{nocharm}$, 
where $X_{nocharm}$ denotes any hadronic 
charmless final state. 
 A measurement of 
the corresponding branching ratio would allow the extraction of
the ratio $|V_{ub}/V_{cb}|$, which is poorly known 
at present  \cite{LenzNierste}. 
At the quark level, there are decay modes with three-body final states,
$b \to q' \overline{q}' q$ ($q'=u,d,s$; $q=d,s$) and the modes
$b \to q g$, with two-body final-state topology.
The component $b \to s g$
of the charmless hadronic decays is expected to manifest itself in 
kaons with high momenta (of order $m_b/2$), owing to its two-body nature
\cite{Rathsman}. 

\subsubsection{$B \rightarrow X_d \gamma$}
With respect to new physics, also 
the $B \rightarrow X_d \gamma$ decay
is of specific interest, because its CKM 
suppression by the factor $|V_{td}|^2/|V_{ts}|^2$ in the SM may 
not be true in extended models.
Moreover, a future measurement of 
the $B \rightarrow X_d \gamma$ decay rate will help to drastically reduce
the currently allowed region
of the CKM-Wolfenstein parameters $\rho$ and $\eta$.

Most of the theoretical improvements carried out in the
context of the decay $B \rightarrow X_s \gamma$ (see sections
\ref{NLLQCDcorrections} and \ref{sectionnonpert})  can 
straightforwardly be adapted
for the decay $B \rightarrow X_d \gamma$. As for the
former decay, the NLL-improved and
power-corrected decay rate for $B \rightarrow X_d \gamma$ has
much reduced theoretical uncertainty, which would allow a more
precise  extraction  of
 the CKM parameters from the measured branching ratio.

The perturbative 
QCD corrections in the decay  $B \rightarrow X_d \gamma$ can be treated 
in complete analogy to the ones in the decay $B \rightarrow X_s \gamma$.
The effective Hamiltonian  is the same 
in  the processes $b \to s \gamma$ and $b \to d \gamma$
up to the obvious
replacement of the $s$-quark field by the $d$-quark field.  
However, as $\lambda_u$ for $b \to d \gamma$ is not small with respect to
$\lambda_t$ and $\lambda_c$, one also has to encounter 
the operators proportional
to $\lambda_u$.
The matching conditions $C_i(m_W)$
and the solutions
of the RG equations, yielding $C_i(\mu_b)$, coincide
with those needed for the process $B \to X_s \gamma$.
The power corrections in $1/m_b^2$ and $1/m_c^2$ (besides the CKM factors)
are also the same for the two modes.

The long-distance 
contributions
from the intermediate $u$-quark in the penguin loops, however, are different.
These are  suppressed in the $B \rightarrow X_s \gamma$ mode by  the 
unfavourable CKM matrix elements. In $B \rightarrow X_d \gamma$, there 
is  no CKM suppression  and one
has to include the long-distance intermediate $u$-quark contributions,
which  can only be modelled at present. However, 
these contributions
 are estimated to be rather small \cite{LDUP}. 
Moreover, it
must be stressed that
there is no spurious enhancement  of the form $\log (m_u/\mu_b)$ 
in the perturbative contribution to the matrix elements
$\bra {X_d}\gamma|O_{iu}|{B} \ket$ ($i=1,2$)
as shown by the explicit calculation in \cite{GHW} and also
discussed in \cite{STERMAN}. In other words, the limit $m_u
\to 0$ can be taken.

The predictions for the 
$ B \rightarrow X_d \gamma$ decay given in \cite{AG7}
show that for $\mu_b=2.5$ GeV (and the central values of the input
 parameters) the difference between the LL and NLL results is 
$\sim 10\%$, increasing the branching ratio in the NLL case.
For a fixed value of the CKM-Wolfenstein parameters 
$\rho$ and $\eta$, the theoretical uncertainty of the branching
ratio is:
\begin{equation}
\Delta {\cal B}(B \rightarrow X_d \gamma)/ {\cal B}(B \rightarrow 
X_d \gamma)   = 
\pm (6-10)\%. 
\end{equation}
Of particular theoretical interest is the ratio of the
branching ratios, defined as
\begin{equation}
\label{dsgamma}
R(d\gamma/s\gamma) \equiv \frac{{\cal B}(B \to X_d \gamma)}
                           {{\cal B}(B \to X_s \gamma)},
\end{equation}
in which a good part of the theoretical uncertainties cancels. 
This suggests that 
a future  measurement of $R(d\gamma/s\gamma)$ will have a large impact on
the CKM phenomenology.

Varying the CKM-Wolfenstein parameters $\rho$ and $\eta$ in the range
$-0.1 \leq \rho \leq 0.4$ and $0.2 \leq \eta \leq 0.46$ and taking into
account other parametric dependences stated above, the 
results (without electroweak corrections) are
\begin{eqnarray}
\label{summarybrasy}
6.0 \times 10^{-6} &\leq &
 {\cal B}(B \rightarrow X_d \gamma)   \leq 2.6 \times 10^{-5}~, \nonumber\\
0.017 &\leq & R(d\gamma/s\gamma) \leq 0.074~.\nonumber
\end{eqnarray}
These quantities are expected to be measurable at the  
high-luminosity $B$ facilities.

\subsubsection{$B \rightarrow X_s l^+l^-$}

\begin{figure}
\begin{center}
\epsfig{figure=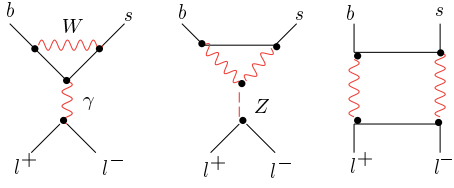,width=6.2cm}
\end{center}
\vspace{-0.5cm}
\caption{One-loop contributions to the decay $B \rightarrow X_s l^+ l^-$.}
\label{llpicture}
\end{figure}

The inclusive $ B \rightarrow X_s l^+l^-$ decay will also be accessible 
at the $B$ factories. In comparison with the 
$B \rightarrow X_s \gamma$  decay, it
presents a complementary and also more complex test of the SM since different
contributions add to the decay rate (fig. \ref{llpicture}).
Because of kinematic observables such as the invariant 
dilepton mass spectrum and 
 the forward-backward asymmetry, 
it is  particularly attractive. It is  also dominated by perturbative 
contributions, if one eliminates $c\bar{c}$ resonances with 
the help of kinematic cuts.

Using heavy quark expansion, nonperturbative corrections scaling
with $1/m_b^2$ and $1/m_c^2$ can be calculated 
quite analogously to  those in the decay $B \rightarrow X_s \gamma$
\cite{Rey}. However, there are also  on-shell $c\bar{c}$ resonances,
which one has to take into account.
While in the decay $B \rightarrow X_s \gamma$ (on-shell photon) 
the intermediate $\psi$ background for example, namely 
${B} \to \psi X_s$ followed by $\psi \to X' \gamma$, is 
suppressed  and can be subtracted 
from the $B \rightarrow X_s \gamma$ decay rate
(see section \ref{experimental}),
the $c\bar{c}$ resonances show up as large peaks in the dilepton 
invariant mass spectrum 
in the decay $B \rightarrow X_s l^+l^-$ (off-shell photon). However,  
these resonances can be removed by appropriate  kinematic cuts
in the invariant mass spectrum:
In the 'perturbative window', namely  
$ 0.05 < \hat{s} = (m_{l^+l^-} / m_b)^2 < 0.25 $,
theoretical predictions for the invariant mass spectrum
are dominated by the purely perturbative contributions, 
and theoretical precision comparable with  the one reached  
in the decay $B \rightarrow X_s \gamma$ is in principle possible.

The present status of the calculation of the perturbative contributions 
is the following: 
the effective Hamiltonian relevant to  
$B \to X_s l^+ l^-$ in the SM reads

\begin{equation}
\label{heffll}
H_{eff}(B \to X_s l^+ l^-)
       = - \frac{4 G_{F}}{\sqrt{2}} \, \lambda_{t} \, \sum_{i=1}^{10}
C_{i}(\mu) \, {\cal O}_i(\mu) \quad ,
\end{equation}

Compared with the decay $B \rightarrow X_s \gamma$ (see (\ref{heff})),  
the effective Hamiltonian~(\ref{heffll}) contains in this case two additional
operators:
 
\begin{equation}
\begin{array}{ll}
{\cal O}_{9\phantom{a}}                 \,= &\!
  \displaystyle{\frac{e^2}{16\pi^2}} \,
 (\bar{s} \gamma_\mu  P_L b)\, (\bar{l} \gamma^\mu l)\,, \\[2.0ex]
{\cal O}_{10}                           \,= &\! 
  \displaystyle{\frac{e^2}{16\pi^2}} \,
 (\bar{s} \gamma_\mu P_L b)\,  (\bar{l} \gamma^\mu \gamma_5 l)\,.    
\end{array}
\label{leptonop}
\end{equation}

It turns out that 
the first large logarithm 
of the form $\log(m_b/M) \quad (M=m_W)$ arises already without 
gluons,
because the operator ${\cal O}_2$ mixes into ${\cal O}_9$ at one loop 
(the pair $c \bar{c}$ in ${\cal O}_2$ can be closed to form a loop,
and an off-shell photon producing a $l \,\bar{l}$ pair can 
be radiated from a quark line). 
This possibility has no 
correspondence in the $B \to X_s \gamma$ case within the SM.
Consequently, the decay amplitude is ordered according to 
\begin{equation}
G_F \,\, \log(m_b/M) \,\,\,\, (\alpha_s(m_b) \, \log(m_b/M))^n \quad (LL),
\nonumber
\end{equation}
\begin{equation}
G_F \,\, \log(m_b/M) \,\, \alpha_s(m_b) \,\,\,\, (\alpha_s(m_b) \, \log(m_b/M))^n 
\quad   (NLL),
\end{equation} 
which should be compared with (\ref{LLQCD}) and (\ref{NLLQCD}). 
To technically achieve the resummation of these terms, it is convenient
to  redefine magnetic, chromomagnetic and lepton-pair operators 
$ {\cal O}_{7}$, $ {\cal O}_{8}$, $ {\cal O}_{9}$, 
and $ {\cal O}_{10}$ and the corresponding coefficients 
as follows~\cite{MM,BurasMuenz}:
\begin{equation}
\label{reshuffle}
{\cal O}_i^{new} = \frac{16 \pi^2}{g_s^2} {\cal O}_i, 
\quad 
C_i^{new} = \frac{g_s^2}{16\pi^2} C_i \quad 
\quad (i=7,...,10). 
\end{equation}
This redefinition
enables one to proceed according to the three 
calculational steps presented in section \ref{strong}, 
when calculating the decay amplitude \cite{MM,BurasMuenz}.
In particular, the one-loop mixing of the operator ${\cal O}_2$ 
with the operator
${\cal O}_9^{new}$ appears formally at order $g_s^2$, after the 
reshufflings in (\ref{reshuffle}).

The QCD calculation up to NLL precision can be found in \cite{MM,BurasMuenz}.
However, the $LL$ term in the series accidentially
turns out to be 
small. In order to reach the same accurary as in the
case of the NLL prediction for $B \rightarrow X_s \gamma$    
one has to include the NNLL order contribution in the 
$B \rightarrow X_s l^+ l^-$ calculation. 

Large parts of the NLL calculation in the decay $B \rightarrow X_s \gamma$, 
reviewed in section \ref{NLLQCDcorrections}, 
can be taken over and used in the NNLL calculation within the decay
$B \rightarrow X_s l^+ l^-$. However, the {\it complete} NNLL enterprise
- following the standard three steps in the formalism of effective theories
(see section \ref{strong}) - 
is a formidable task:

\begin{itemize} 

\item Step 1: \,
In \cite{MISIAKBOBETH} the complete Step 1 up to NNLL precision was presented. 
The authors did the  two-loop matching 
for all the operators relevant to $B \rightarrow X_s l^+l^-$ 
(including a confirmation of the $B \rightarrow X_s \gamma$ NLL matching 
results of \cite{Adel,GH,CDGG,INFRARED}).
The inclusion of this  NNLL contribution already removes the large
matching scale ($\mu_W$) uncertainty of around $16 \%$ present in the 
NLL prediction
of $B \rightarrow X_s l^+l^-$.
As usual the partonic decay width is normalized by the semileptonic
decay width in order to get rid of uncertainties due to the fifth
power in $m_b$: 
\begin{equation}
R^{l^+l^-}_{quark}(\hat{s})=
\frac{1}{\Gamma(b \to X_c e\bar{\nu})}\frac{d\Gamma(b\to
X_s l^+ l^-)}{d\hat{s}} \,\,.
\end{equation}

One finds the following partial NNLL
prediction \cite{MISIAKBOBETH}:
\begin{eqnarray}
&& {\cal B}(B \rightarrow X_s l^+ l^-)_{Cut:\,\,\hat{s}  
\in [0.05,0.25]} = \nonumber\\
&=&\, {\cal  B}(B \rightarrow X_c e \overline{\nu}) \, \int_{0.05}^{0.25}  \, d \hat{s} \, \big[ R^{l^+ l^-}_{quark}(\hat{s}) + \delta_{1/m_b^2} R(\hat{s}) +
\delta_{1/m_c^2} R(\hat{s})  \big]
\nonumber\\
&=&\, 0.104 \, [ (1.36 \pm 0.18_{scale} \,) + 0.06 - 0.02] \, 10^{-5} \nonumber\\
&=&\, ( 1.46  \pm 0.19_{scale} \,) \, 10^{-6}
\label{prefinal}
\end{eqnarray}

$\delta_{1/m_b^2} R(\hat{s})$ and $\delta_{1/m_b^2} R(\hat{s})$ 
are the nonperturbative contributions discussed in section 
\ref{sectionnonpert}. 
The quoted  error in (\ref{prefinal}) 
reflects only the $\mu_b$ scale uncertainty. 
This purely perturbative uncertainty should get significantly 
reduced by contributions within Step 3 of the NNLL program,
namely  the two-loop QCD corrections to the matrix element 
of the four-quark operators ${\cal O}_2$. 
The error due to the uncertainties in the input parameters
and to other contributions was not estimated  in
\cite{MISIAKBOBETH},
at this intermediate stage of the NNLL calculation.

\item Step 2:\, The most important NNLL contribution from
 the three-loop renormalization group
evolution of the Wilson coefficients from the matching scale $\mu_W$ 
to the low scale $\mu_b$, namely the 
 three-loop anomalous 
dimensions corresponding to the mixing of the four-quark operators 
${\cal O}_i  \quad(i=1...6)$ into the dipole operators  ${\cal O}_7$ and 
${\cal O}_8$, can be taken over from the NLL calculation in the decay 
$B \rightarrow X_s \gamma$ \cite{Mikolaj}. However, 
the analogous three-loop anomalous dimensions corresponding to the 
mixing of the four-quark operators into the operator ${\cal O}_9$
is missing. In \cite{MISIAKBOBETH} an estimate was made which suggests 
that the numerical influence of these missing NNLL
contributions to the branching ratio of $B \rightarrow X_s l^+ l^-$ 
is small.

\item Step 3: \, Within the NLL $B \rightarrow X_s \gamma$ calculation the 
two-loop matrix elements of the four-quark operator ${\cal O}_2$
for an on-shell photon were calculated in \cite{GHW}
and quite recently confirmed in \cite{Burasnew}. This calculation 
was extended to the case of an off-shell photon
\cite{Christophnew},
which corresponds to a NNLL contribution relevant to the decay
$B \rightarrow X_s l^+l^-$. 
The calculation includes also that part of the corresponding 
gluon bremsstrahlung which is needed to cancel infrared and collinear
singularities of the virtual corrections.
If one includes also this  NNLL piece in the partonic
NNLL prediction for the decay $B \rightarrow l^+l^-$, one gets
\cite{Christophnew}
\begin{equation}
\int_{0.05}^{0.25}  \, d \hat{s} \,  R^{l^+ l^-}_{quark}(\hat{s}) \, = \, (1.25 \pm 0.08_{scale} \,)\, \times  10^{-5} 
\label{final}
\end{equation}
Again the only error given corresponds to the uncertainty of the low  scale
$\mu_b$. As expected the inclusion of the two-loop virtual corections to the
four-quark operator ${\cal O}_2$ has reduced this scale ambiguity 
from $\pm 13\%$ down to $\pm 6.5\%$. The authors of \cite{Christophnew}
also analyse the error due to the uncertainty in the input parameter
$m_c/m_b$ and find an uncertainty of $\pm 7.6\%$ within the 
partonic quantity.

Within the Step 3 of the NNLL calculation,  the renormalization
group invariant two-loop matrix element of the operator ${\cal O}_9$
is not calculated yet.  Because this contribution includes  no logarithms, 
the scale dependence of the NLL prediction is not sensitive to this
NNLL contribution.

\end{itemize}

\begin{figure}
\begin{center}
\epsfig{figure=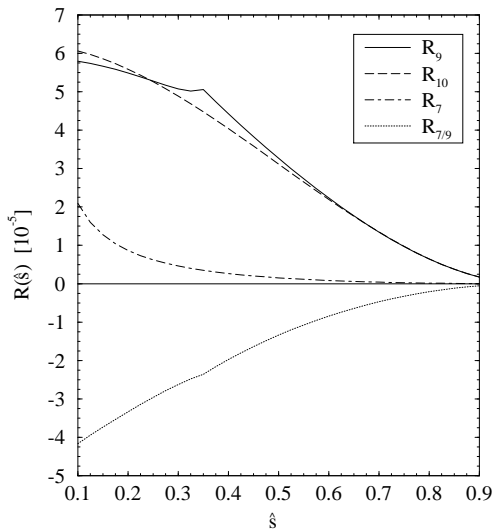,width=6.5cm}
\end{center}
\vspace{-0.5cm}
\caption{Comparison of the different short-distance contributions to $R_{quark}(\hat s)$ (NLL precision), from \cite{BurasMuenz}.}
\label{Rpartiell}
\end{figure}

One could think that within this perturbative window at  low 
$\hat{s} \in [0.05,0.25]$, 
one is only sensitive to $C_7$ which would be redundant
information, since we already know it 
from the decay $B \rightarrow X_s \gamma$. 
However, as was explicitly shown  in \cite{MM,BurasMuenz}, one is also sensitive 
to  the new Wilson coefficients $C_9$ and $C_{10}$ and interference terms 
in the low $\hat{s}$ regime with $\hat{s} = m_{l^+l^-}/m_b^2 \in [0.05,0.25]$
(see fig. \ref{Rpartiell} where the various perturbative contributions
to $R_{quark}$ (with NLL precision) are plotted).

Together with the decay $B \rightarrow X_s \gamma$, 
the inclusive $B \rightarrow X_s l^+ l^-$ decay  will make 
precision flavour physics possible, if one can also measure 
the kinematic  variables in the $B \rightarrow X_s l^+l^-$ decay 
precisely. 
As was first  advocated in \cite{AliMannel}, 
\begin{itemize}
\item the invariant dilepton mass spectrum 
\begin{equation}
d \Gamma ( B \rightarrow X_s l^+ l^- ) \, / \,  d \hat{s},
\end{equation}
\item the forward-backward charge asymmetry 
\begin{equation} 
A(s) = \int_{-1}^{1} \, d \mbox{cos}\theta \, \, 
d^2 \Gamma ( B \rightarrow X_s l^+l^- ) \, / \, d \hat{s} \, d \mbox{cos} \theta \,\,\, 
\mbox{sgn}(\mbox{cos}\theta)
\end{equation}
\item
and the decay rate of $B \rightarrow X_s \gamma$, 
\begin{equation}
\Gamma(B \rightarrow X_s \gamma)
\end{equation}
\end{itemize}
determine the magnitude and also the sign 
of the three Wilson coefficients $C_7,C_8,$ and $C_{10}$, and allow for a 
model-independent analysis of rare $B$ decays.
For the measurements of these kinematic distributions, however,
high statistics will be necessary.

\setcounter{equation}{0}
\section{Indirect search for supersymmetry}

Today supersymmetric models are given priority in our search for 
new physics beyond the SM. This is primarily suggested by theoretical 
arguments related to the well-known hierarchy problem. 
The decay  $B \rightarrow X_s \gamma$ is sensitive to the 
mechanism of supersymmetry breaking because 
in the limit of exact supersymmetry, the decay rate would
be just zero:
\begin{equation}
{\cal B}(B \to X_s \gamma) = 0.  
\end{equation}
This follows from an argument first given by Ferrara and Remiddi in 1974
\cite{Ferrara}. 
In that work  the absence of the anomalous 
magnetic moment in a supersymmetric abelian gauge theory was shown.
The necessary mechanism of supersymmetry breaking, however, is unknown and 
leads to a proliferation of free parameters 
in the (unconstrained)  minimal 
supersymmetric standard model (MSSM).

There are two types of new contributions to flavour changing
neutral currents in the MSSM: 
CKM-induced contributions, which are induced by a charged 
Higgs or a chargino, and generic new  contributions, which are induced by 
flavour mixing  in the squark-mass matrix.
The structure of the MSSM does not explain the suppression of flavour
changing neutral currents 
which is observed in experiments.
This is the essence of the well-known supersymmetric flavour problem.

In the framework of the MSSM there are at present three favoured concrete
supersymmetric models. They solve the supersymmetric 
flavour problem by a specific mechanism through which the sector of 
supersymmetry breaking 
communicates with the sector accessible to experiments: in the 
minimal supergravity model (mSUGRA)~\cite{MSUGRA},  
 supergravity is the corresponding
mediator; in the other two models this role is fulfilled by gauge 
interactions (GMSB)~\cite{GMSB}
 and by anomalies (AMSB)~\cite{AMSB}. 
Furthermore, there are other 
classes of models in which the flavour problem is solved by particular 
flavour symmetries~\cite{FLAVOUR}.

Flavour violation thus originates
from the interplay  between the dynamics of flavour and the mechanism of  
supersymmetry breaking.
The model-independent analysis of rare $B$ and $K$ decays  
therefore can contribute to the question of 
which mechanism ultimately breaks the supersymmetry and will
thus yield important (indirect) information on the construction 
of supersymmetric extensions of the SM.
In this context it is important to analyse the correlations between the
different information from rare $B$ and $K$ decays.

As was already emphasized in the introduction, inclusive rare 
decays, as loop-induced processes, are particularly 
sensitive to new physics and theoretically clean.
Neutral flavour transitions involving third-generation quarks,
typically in the $B$ system, do not pose yet serious threats to specific
models. However, 
despite the relatively large experimental uncertainties, 
the rare decay $B \rightarrow X_s \gamma$
has already carved out some regions in the space of free
parameters of most of the models in the classes mentioned above
 (see \cite{THEO},\cite{TUNED} and references therein). 
Once more precise data from the $B$ factories are  available, 
this decay will undoubtedly gain efficiency
in selecting the viable regions of the parameter space in the various
classes of models; this may help in discriminating between  the models by
then proposed.
In view of this, it is important to calculate the rate of this 
decay with theoretical uncertainties reduced 
as much as possible, and general enough for generic 
supersymmetric models.

While in the SM, the rate for $B \to X_s \gamma$
is known up to NLL in QCD, 
the calculation of this decay rate within supersymmetric models 
is still far from this level of sophistication.
There are several contributions to the decay amplitude: 
besides the 
$W\,t$-quark and the $H\,t$-quark contributions, there are also
the chargino, gluino and neutralino contributions.
In most of the phenomenological analyses of the decay 
$B \rightarrow X_s \gamma$ these nonstandard contributions   
were {not} investigated  with NLL precision  
as the SM contribution.  
However, as has already been pointed out,
the step from the LL to the NLL precision is also necessary 
in order to check  the validity of the perturbative 
approach in the model under consideration. It is possible that the 
restriction of the parameter space of nonstandard  models, based on 
an LL analysis only, proves to be invalid after the NLL analysis 
is completed. 
Moreover, it was already shown in specific new physics scenarios 
that bounds on the parameter space of  nonstandard 
models are rather sensitive to NLL contributions (see below).

\begin{figure}
\begin{center}
\epsfig{figure=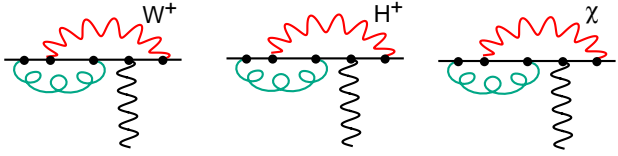,width=8.5cm}
\end{center}
\vspace{-0.5cm}
\caption{SM, charged Higgs and chargino contribution at the matching scale.}
\label{susycalculation}
\end{figure}

Nevertheless, within supersymmetric models partial 
NLL results are available. 
The gluonic NLL two-loop matching contributions were 
recently presented  \cite{mikolaj99}.  
A complete NLL calculation
of the $B \rightarrow X_s \gamma$ branching ratio in the simplest
extension of the SM, namely the Two-Higgs-Doublet Model (2HDM), is
already available \cite{CDGG,BG}. 
In the 2HDM of Type II (which already represents 
a good approximation for 
gauge-mediated
supersymmetric models with large $\tan\beta$ where the charged Higgs
contribution dominates the chargino contribution), 
the $B \rightarrow X_s \gamma$ is only sensitive to two
parameters of this model, the mass of the charged Higgs boson 
and $tan \beta$. 
Thus, the  experimental data of 
the decay $B \rightarrow X_s \gamma$ allows for stringent bounds on these
two  parameters which are much more restrictive 
than the lower bound on the charged Higgs mass found in the direct 
search at LEP.  
One also finds that these
indirect bounds are very sensitive to NLL QCD corrections and  even
to the two-loop electroweak contributions (see \cite{CDGG,BG}). 

In \cite{guidice}  
a specific  supersymmetric scenario is presented, 
where in particular the possibility of 
destructive interference of the chargino 
and the charged Higgs contribution is analysed. 
The analysis has been  done under two assumptions.
First it is assumed   
that the only source of flavour violation 
at the electroweak scale is that of the SM, 
encoded in the CKM matrix
(minimal flavour violation).
Therefore, the analysis  applies to mSUGRA, GMSB and
AMSB models (in which the same features are assumed  at the 
messenger scale) only when the sources of flavour violation,  
generated radiatively between the supersymmetry breaking scale and 
the electroweak scale, can be neglected with respect to those induced
by the CKM matrix. 
The second assumption is that there exists  
a specific mass hierarchy, in particular 
the heavy gluino limit.
Indeed, the NLL calculation has been done in the limit 
\begin{equation}
\mu_{\tilde{g}} 
\sim O(m_{\tilde{g}},m_{\bar{q}},m_{\tilde{t}_1}) \gg 
\mu_W \sim  O(m_{W},m_{H^+},m_{t},m_{\chi},m_{\tilde{t}_2}).
\label{masshierachy}
\end{equation} 
The mass scale of the 
charginos ($\chi$), and of the lighter stop ($\tilde t_2$) is
the ordinary electroweak scale $\mu_W$, while
the scale $\mu_g$ is 
characteristic of all other strongly interacting supersymmetric
particles (squarks and gluinos) and is 
assumed to be of the order of $1$ TeV. 
NLL QCD corrections have been  calculated up to first order
in  $\mu_W/\mu_{\tilde{g}}$ including the important 
nondecoupling effects \cite{guidice}.

At the electroweak scale $\mu_W$, the new contributions do not induce any new 
operators in this scenario. Thus, the only  step in the new NLL calculation 
beyond the one within the SM is Step 1, 
the matching calculation at the scale
$\mu_W$ where we encounter the two new CKM-induced contributions
of the charged Higgs and the chargino
(see fig. \ref{susycalculation}):
\begin{equation}
 C_{{NLL}} (\mu_W) = C_{{NLL}}^{{SM}} (\mu_W) + 
C_{{NLL}}^{{H^+}} (\mu_W) + C_{{NLL}}^{{\chi}} (\mu_W). 
\end{equation}

\begin{figure}
\begin{center}
\epsfig{figure=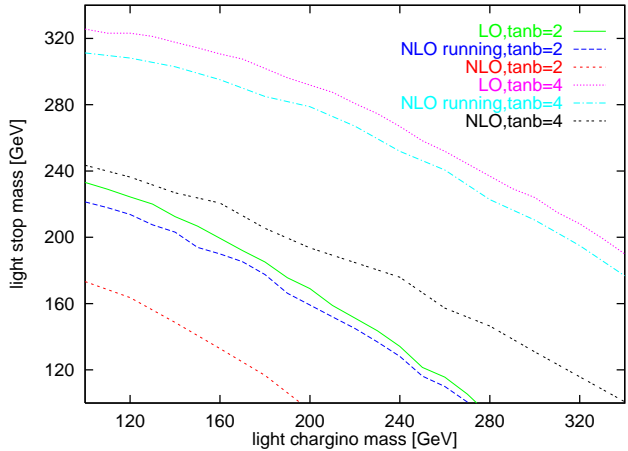,width=9cm}
\end{center}
\caption{
Upper bounds on the lighter chargino and stop masses 
from $B \rightarrow X_s \gamma$ data  in the scenario
(\ref{masshierachy}) if a light charged Higgs mass  
is assumed;
for  $\tan\beta=2$ (three lower curves) and $4$ 
(three upper plots)
the LL, NLL-running and NLL results
(from the top  to the bottom) are shown (see text),
from \cite{guidice}.}
\label{paolo}
\end{figure}

One finds \cite{guidice} 
that,  in this  specific supersymmetric scenario,  
bounds on the parameter space are rather sensitive 
to NLL contributions 
and they lead to a significant reduction of the stop-chargino 
mass region where the supersymmetric contribution has
a large destructive interference with the charged-Higgs boson 
contribution. In fig. \ref{paolo} the 
upper bounds on the lighter chargino and stop masses 
from $B \rightarrow X_s \gamma$ data  in the scenario of
(\ref{masshierachy}) are illustrated if a light charged Higgs mass 
of $m_{H^\pm}=100$ GeV is assumed.
The stop mixing is set to $|\theta_{\tilde t}|<\pi/10$ which 
corresponds to the assumption of a mainly right-handed light stop.
Moreover, $|\mu|<500$ GeV and all heavy masses are around 1 TeV. 
For  $\tan\beta=2$  and $4$ 
the results of the LL, `NLL running' and NLL
calculations are given. The result of
neglecting the new NLL supersymmetric contributions to the 
Wilson coefficients is labelled as "NLL running" and illustrates 
the importance of the NLL chargino contribution \cite{guidice}.

Quite recently, this minimal flavour violation 
scenario was refined and extended to the 
large $tan \beta$ regime by the resummations of terms of the form 
$\alpha_s^n \, tan^{n+1} \beta$ \cite{Carena,Degrassinew}.
The stability of the renormalization group improved perturbation theory 
was reassured for this specific scenario: the resummed NLL results 
in the large $tan \beta$ regime show constraints similar to the
LL results (see also \cite{Deboernew}). 

For example, it is a well-known feature in the mSUGRA model,
that 
depending on the sign of $A_t \cdot \mu$ (where $A_t$ denotes 
the stop mixing parameter) the chargino contribution 
can  interfere constructively ($A_t \cdot \mu > 0$)
or destructively ($A_t \cdot  \mu <  0$)
with the SM and the charged Higgs contribution. 
Therefore, the scenario
$A_t \cdot \mu > 0$ within this model 
requires very heavy superpartners in order to
accommodate the $B \rightarrow X_s \gamma$ data. But also
the case $A_t \cdot \mu <  0$ is constrained in the large $tan \beta$ 
regime where the chargino contribution is strongly enhanced
(for details see \cite{Carena,Degrassinew,Deboernew}).

However, all these NLL analyses  are valid only in 
the heavy gluino regime. 
Thus, these  calculations
cannot be used in particular directions of the parameter 
space of the above listed models in which quantum effects induce a 
gluino contribution as large as the chargino or the SM contributions. 
Nor can it be used as a model-discriminator tool, able to constrain 
the potentially
large sources of flavour violation typical of generic 
supersymmetric models. A complete NLL calculation should also include 
contributions where the gluon is replaced by the  gluino.

The flavour nondiagonal vertex gluino-quark-squark induced by
the flavour violating scalar mass term and trilinear terms
is particularly interesting. This is generically assumed to induce the 
dominant contribution to quark flavour transitions, as this vertex is weighted
by the strong coupling constant $g_s$.
Therefore, it is often taken as the 
only contribution 
to these transitions and in particular to the $B \rightarrow X_s \gamma$
decay, when attempting to obtain order-of-magnitude upper bounds
on flavour violating terms in the scalar potential \cite{MAS,HAG}.
Once the constraints coming from the experimental measurements are imposed, 
however, the gluino contribution is reduced to values such that the SM 
and the other supersymmetric contributions can no longer  be neglected. 
Any LL and NLL calculation of the $B \rightarrow X_s \gamma$
rate in generic supersymmetric models, therefore, should then include
all possible contributions.

\begin{figure}
\begin{center}
\epsfig{figure=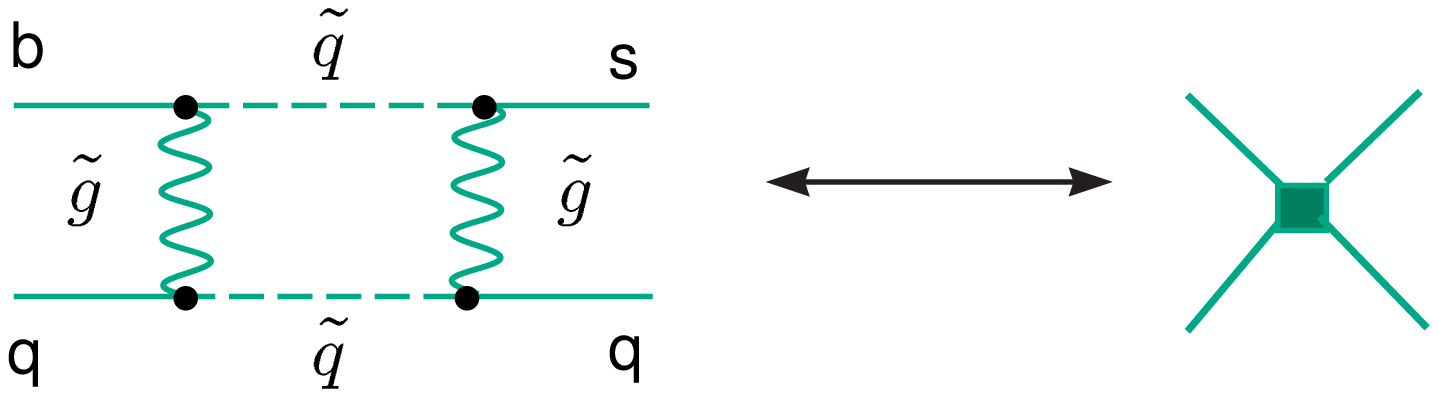,width=7cm}
\hspace{1cm}
\epsfig{figure=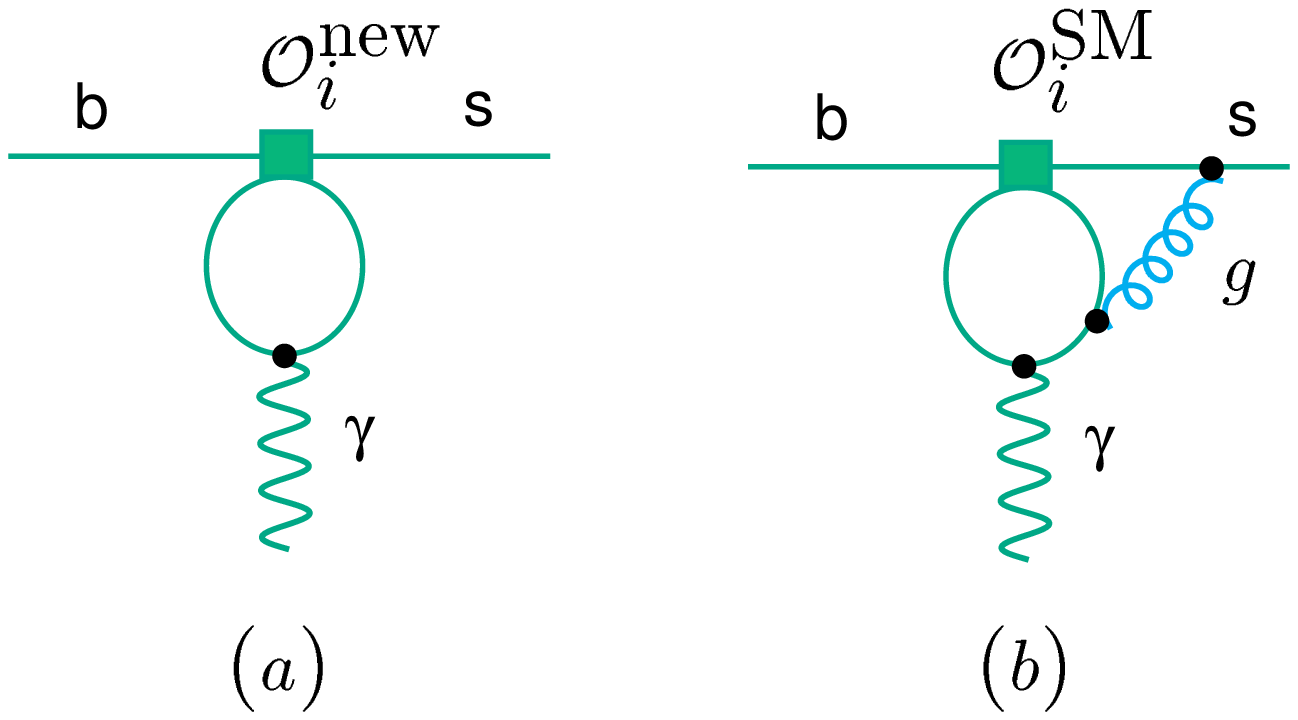,width=5cm}
\vspace{-0.5cm}
\end{center}
\caption{
Matching of gluino-squark box on new scalar operators (left frame);
mixing of new (scalar) operators at one-loop (a) in constrast to the 
vectorial operators of the SM (b) who mix at two-loop only (right frame).} 
\label{mixingsusy}
\end{figure}

The gluino contribution, however, presents some peculiar features related
to the implementation of the QCD corrections.
In ref. \cite{OUR} this contribution to the decay $B \rightarrow X_s \gamma$
has been  investigated in great detail for
 supersymmetric models with generic soft terms.
The gluino-induced contributions to the decay amplitude for $B \to X_s \gamma$
are of the following form:
\begin{equation}
\alpha_s(m_b) \, (\alpha_s(m_b) \log(m_b/M))^n \, \quad (LL),
\end{equation}
\begin{equation} 
\alpha^2_s(m_b) \,  (\alpha_s(m_b) \log(m_b/M))^n \, \quad (NLL).
\end{equation}
In the matching calculation, 
all factors $\alpha_s$,  regardless of their source, should be 
expressed in terms of the $\alpha_s$ running with five flavours.
In \cite{OUR} it is shown
that 
the relevant operator basis of the SM effective Hamiltonian gets enlarged 
to contain magnetic and chromomagnetic operators with an extra factor of 
$\alpha_s$
 and weighted by a quark mass $m_b$ or $m_c$, and also 
magnetic and chromomagnetic operators of lower dimensionality
where the (small) factor $m_b$ is replaced by the gluino mass.
Furthermore, one finds that  
gluino-squark boxes induce new scalar and tensorial 
four-quark operators, 
which are shown to 
mix into the magnetic operators without gluons.  
On the other hand, the vectorial four-quark  operators 
mix only with an
additional gluon into magnetic ones (fig. \ref{mixingsusy}). 
Thus, they will contribute at the NLL order only.
However, from the numerical point
of view the contributions of the vectorial operators (although NLL) are
not necessarily suppressed w.r.t. the new four-quark contributions;
this is due to the expectation 
that the flavour-violation parameters
present in the Wilson coefficients of the new operators are expected
to be much smaller (or much more stringently constrained)
than the corresponding ones in the coefficients of the vectorial 
operators. This feature shows that a complete
order calculation is important.

In ref. \cite{OUR} the effects of the 
LL QCD corrections on constraints on supersymmetric sources of 
flavour violation are analysed.
To understand the sources of flavour violation that may be present in
supersymmetric models in addition to those enclosed in the CKM matrix,
one has to consider the contributions to the squark mass matrices
\begin{equation}
{\cal M}_{f}^2 =  
\left( \begin{array}{cc}
  m^2_{f,LL}   & m^2_{f,LR} \\
  m^2_{f,RL}  &  m^2_{f,RR}                 
 \end{array} \right) +
\label{squarku}
\nonumber
\end{equation}
\begin{equation}
  \left( \begin{array}{cc}
  F_{f,LL} +D_{f,LL} &  F_{f,LR} \\
 F_{f,RL} & F_{f,RR} +D_{f,RR}                
 \end{array} \right) \quad ,
\nonumber
\label{squarku2}
\end{equation}
where $f$ stands for up- or down-type squarks.
In the super  CKM basis where the quark mass matrices are diagonal 
and the squarks are rotated in parallel to their superpartners,
the $F$ terms  from the superpotential and the $D$ terms 
turn out to be diagonal 
$3 \times 3$ submatrices of the 
$6 \times 6$
mass matrices ${\cal M}^2_f$. This is in general not true 
for the additional terms (\ref{squarku}), originating from  the soft 
supersymmetric breaking potential. 
Because all neutral gaugino couplings are flavour diagonal
in the super CKM basis, the 
gluino contributions to the
decay $b \to s \gamma$ are induced by the off-diagonal
elements of the soft terms 
$m^2_{f,LL}$, $m^2_{f,RR}$, $m^2_{f,RL}$.

\begin{figure}
\begin{center}
\epsfig{figure=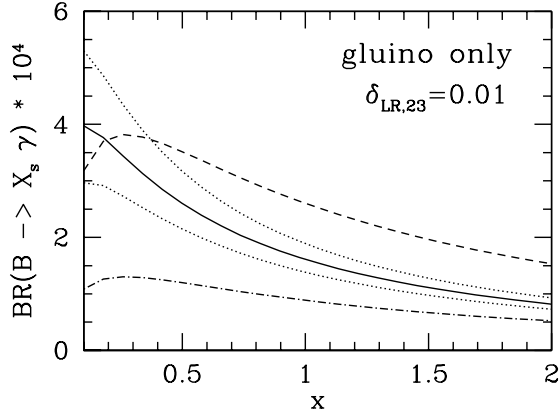,width=8cm}
\end{center}
\vspace{-1.1cm}
\caption[f1]{Gluino-induced branching ratio ${\cal B}(B \to X_s \gamma)$ 
 as a function of $x= m^2_{\tilde{g}}/m^2_{\tilde{q}}$, obtained when
 the only source of flavour violation is $\delta_{LR,23}$ (see text).
}
\label{sizeqcd23lr}
 \end{figure}

It is convenient to select one possible 
source of flavour violation in the squark sector at a time and
assume that all the remaining ones are vanishing. 
It should be stressed that one already excludes any kind
of interference effects between different sources of flavour
violation in this way. 
Following
ref.~\cite{MAS}, all diagonal entries in 
$m^2_{\,d,\,LL}$, $m^2_{\,d,\,RR}$, and $m^2_{\,u,\,RR}$
are set to be equal and their common value is denoted by
$m_{\tilde{q}}^2$.  The branching ratio can then be studied as a
function of 

\begin{equation} 
\delta_{LL,ij} = \frac{(m^2_{\,d,\,LL})_{ij}}{m^2_{\tilde{q}}}\,, 
\hspace{0.1truecm} \qquad
\delta_{RR,ij} = \frac{(m^2_{\,d,\,RR})_{ij}}{m^2_{\tilde{q}}}\,, 
\hspace{0.1truecm} 
\label{deltadefa}
\end{equation}

\begin{equation} 
\delta_{LR,ij} = \frac{(m^2_{\,d,\,LR})_{ij}}{m^2_{\tilde{q}}}\, 
\, (i \ne j).
\label{deltadefb}
\end{equation}

\begin{figure}
\begin{center}
\epsfig{figure=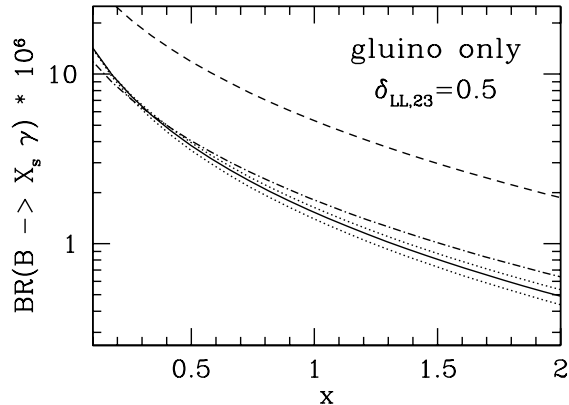,width=8cm}
\end{center}
\vspace{-1.1cm}
\caption[f1]{
Gluino-induced branching ratio ${\cal B}(B \to X_s \gamma)$ 
 as a function of $x= m^2_{\tilde{g}}/m^2_{\tilde{q}}$, obtained when
 the only source of flavour violation is $\delta_{LL,23}$ (see text).}
\label{sizeqcd23ll}
\end{figure}

The remaining crucial parameter needed to determine the 
branching ratio is $x = m^2_{\tilde{g}}/m^2_{\tilde{q}}$,
where $m_{\tilde{g}}$ is the gluino mass.
Figures~\ref{sizeqcd23lr} and~\ref{sizeqcd23ll}
show the LL QCD corrections to the 
gluino contribution.

In these figures
the solid lines show  the QCD corrected
branching ratio, when only $\delta_{LR,23}$ or
$\delta_{LL,23}$ are nonvanishing.  
The branching ratio is plotted as a function of
$x$, using
$m_{\tilde{q}}=500\,$GeV.  The  dotted lines show the range of variation
of the branching ratio, when the
renormalization scale $\mu$ varies in the interval $2.4$--$9.6\,$GeV. 
Numerically, the scale uncertainty in
${\cal B}(B \to X_s \gamma)$ is about
$\pm 25\%$. An extraction of bounds on the $\delta$
quantities more precise than just an order of magnitude or less,
would, therefore, require the inclusion of NLL QCD
corrections. It should be noticed, however, that the inclusion of the
LL QCD corrections has already removed the large ambiguity on the
value to be assigned to the factor $\alpha_s(\mu)$ in the
gluino-induced operators. Before
adding QCD corrections, the scale in this factor can assume all values
from $O(m_b)$ to $O(m_W)$: the difference between the value 
of  ${\cal B}(B \to X_s \gamma)$
obtained when taking $\alpha_s(m_b)$ and that obtained when taking
 $\alpha_s(m_W)$ is
of the same order as the LL QCD corrections.  The corresponding values
of  ${\cal B}(B \to X_s \gamma)$ for the two extreme choices of $\mu$ are
indicated in Figs.~\ref{sizeqcd23lr} and~\ref{sizeqcd23ll} by
the dot-dashed lines ($\mu=m_W$) and the dashed lines
($\mu=m_b$). The
choice $\mu = m_W$ gives values for the non-QCD corrected 
${\cal B}(B \to X_s \gamma)$ relatively close to the band obtained when the
LL QCD corrections are included, if only $\delta_{LL,23}$ is
nonvanishing. Finding a corresponding value of $\mu$ that minimizes
the QCD corrections in the case studied in Fig.~\ref{sizeqcd23lr},
when only $\delta_{LR,23}$ is different from zero,
depends strongly on the value of $x$.
In the context of the full LL result, it is important
to stress that the explicit $\alpha_s$ factor 
                       has to be evaluated 
- like the Wilson coefficients - at a scale $\mu=O(m_b)$.

In spite of the large uncertainties still affecting the branching ratio 
${\cal B}(B \to X_s \gamma)$ at  LL in QCD, it is possible
to extract indications of the size that the $\delta$-quantities 
may maximal acquire without inducing conflicts with the 
experimental measurements (see \cite{OUR}). 

Finally, it should be emphasized that a consistent analysis 
of the bounds on the sfermion mass matrix should also  include 
 interference effects between the various contributions.
For this issue we refer to a quite recent paper    \cite{NEWNEW}, 
where  
the interplay  between the various 
sources of flavour violation and the interference effects 
of SM, gluino,
chargino, neutralino and charged Higgs boson contributions is
systematically analysed. 
New bounds on  simple combinations of
elements of the soft part of the squark mass matrices
are found to be, in general, one order of magnitude weaker 
than the bound on the single off-diagonal element $\delta_{LR,23}$, which 
was derived in previous work \cite{MAS,Masiero2001} by
neglecting any kind of interference effects.

\setcounter{equation}{0}
\section{Direct CP violation in $b/s$ transitions}
\label{CPsection}

Detailed measurements
of CP asymmetries in rare $B$ decays will be possible in the near 
future. Theoretical predictions for the {\it normalized} CP asymmetries of
the  inclusive channels (see~\cite{AG7,KaganNeubert,SoniWu}) within 
the SM lead to 
\begin{equation}
\label{SMpredict} 
\nonumber 
\alpha_{CP}({B \rightarrow  X_{s/d} \, \gamma}) =
\frac{\Gamma(\bar B \rightarrow X_{s/d}\gamma)
     -\Gamma(B \rightarrow  X_{\bar s/\bar d}\gamma)}
     {\Gamma(\bar B \rightarrow  X_{s/d} \gamma)
     +\Gamma(B \rightarrow  X_{\bar s/\bar d}\gamma)}
\end{equation}
\begin{equation}                    
  \alpha_{CP}({B \rightarrow  X_s \gamma}) \approx 0.6 \%, \qquad 
  \alpha_{CP}({B \rightarrow B_d \gamma}) \approx  -16 \%
\label{SMnumbers}
\end{equation}
when the best-fit values for the CKM parameters \cite{CKMfit} 
are used.

The leading partonic contribution to the CP asymmetries  
is given by 
\begin{eqnarray}
\label{SMCPexplicit}
\alpha_{CP}({B \rightarrow  X_{s/d} \, \gamma}) \simeq  
\frac{10^{-2}}{|C_7|^2} ( 1.17 \times {\rm Im}\left[C_2C_7^*\right] 
                - 9.51 \times {\rm Im}\left[C_8C_7^*\right] 
               \\
+0.12\times {\rm Im}\left[C_2C_8^*\right] 
-9.40 \times {\rm Im}\left[\epsilon_{s(d)}C_2\left(
                C_7^*-0.013 \; C_8^*\right)\right] ) \nonumber;\\
\epsilon_s = \frac{V_{us}^*V_{ub}}{V_{ts}^*V_{tb}} \simeq 
-\lambda^2(\rho-i\eta), \quad \quad   
\epsilon_d = \frac{V_{ud}^*V_{ub}}{V_{td}^*V_{tb}} \simeq
        \frac{\rho -i\eta}{1-\rho+i\eta} .
\nonumber
\end{eqnarray}

The large coefficient of the second term in (\ref{SMCPexplicit})
has triggered an attractive scenario in which an enhanced 
chromomagnetic dipole contribution, $C_8$, induces a large direct 
CP violation in the decay $B \rightarrow X_s \gamma$.
Such a possible enhancement 
of the chromomagnetic contribution would lead to a natural explanation
of the phenomenology of semileptonic 
$B$ decays  and also of charm production in 
$B$ decays \cite{KaganNeubert}.

An analysis for the leptonic counterparts  is presented in
\cite{Hiller}.
The normalized CP asymmetries may also be calculated for
exclusive decays; however, these results are model-dependent. An
example of such a calculation may be found in \cite{GSW}.

CLEO has already presented a measurement of the CP asymmetry in
inclusive $b \to s \gamma$ decays, yielding \cite{CleoCP}
\begin{equation}
\alpha_{CP}(B \rightarrow  X_s \gamma) = 
(-0.079 \pm 0.108 \pm 0.022) \times (1.0 \pm 0.030) \, , 
\end{equation}
which indicates that very large effects are already excluded.

Supersymmetric predictions for the CP asymmetries in 
$B \rightarrow X_{s/d} \gamma$ depend strongly on what is
assumed for the supersymmetry-breaking sector and are, thus,  
a rather model-dependent issue. 
The minimal supergravity model cannot account for large 
CP asymmetries beyond $2\%$ because of the constraints coming 
from the electron and neutron electric dipole moments
\cite{Goto}. However, more general models 
allow for larger asymmetries,  of the order of $10 \%$ 
or even larger \cite{Aoki,KaganNeubert}.
Recent studies of the $B \to X_d \gamma$ rate asymmetry  
in specific models led to asymmetries between $-40 \% $ and $+40\%$
\cite{Recksiegel} or  $-45 \%$ and $+ 21 \%$ \cite{Asatrian}.
In general, CP asymmetries may lead to clean 
evidence for  new physics by a significant deviation from the SM
prediction.  From (\ref{SMnumbers}),  it is obvious  that a large 
CP asymmetry  in the 
$B \rightarrow X_s \gamma$ channel or a positive CP asymmetry in the 
inclusive $B \rightarrow X_d \gamma$ channel would be a clear signal
for new physics. 

In \cite{mannelhurth} it was pointed out that  
the exclusive and inclusive decays of the form $b \to s \gamma$ and
$b \to d \gamma$, as well as their leptonic counterparts, provide a
stringent test, if the CKM matrix is indeed the only source of
CP violation. 
Using  U-spin, which is the $SU(2)$ subgroup of flavour $SU(3)$ relating
the $s$ and the $d$ quark and which is already a well-known tool 
in the context of nonleptonic decays \cite{Fleischer:1999pa,GronauBCP4},
one derives  relations between the CP asymmetries of the exclusive 
channels $B^- \to K^{*-} \gamma$ and $B^- \to \rho^- \gamma$ and 
of the inclusive channels $B \rightarrow X_s \gamma$ and 
$B \rightarrow X_d \gamma$.
One should make use  of the U-spin symmetry only
with respect to the strong interactions.
Moreover, within exclusive final states,
the vector mesons like the U-spin doublet ($K^{*-}$, $\rho^-$) 
are favoured as final states because 
these have masses much larger than the (current-quark) masses
of any of the light quarks. Thus one expects, for the ground-state 
vector mesons, the U-spin symmetry to be quite accurate in spite of the
nondegeneracy of $m_d$ and $m_s$. 
Defining the rate asymmetries ({not} the {\it normalized} CP asymmetries)
by  
\begin{equation} \label{ratediff}
\Delta \Gamma (B^- \to V^- \gamma) = \Gamma (B^- \to V^- \gamma) - \Gamma (B^+ \to V^+ \gamma)
\end{equation}
one arrives at the following relation \cite{mannelhurth}:
\begin{equation} 
 \Delta \Gamma (B^- \to K^{*-} \gamma) +
\Delta \Gamma (B^- \to \rho^- \gamma) = b_{exc} \Delta_{exc}
\end{equation}
where the right-hand side is written as a product of a relative 
U-spin breaking $b_{exc}$ and a typical size  $\Delta_{exc}$ 
of the CP violating rate difference. 
In order to give an estimate of the right-hand side, one can use
the model result  from \cite{GSW}  for $\Delta_{exc}$,
\begin{equation}
\Delta_{exc}  = 2.5 \times 10^{-7}\,\,  \Gamma_B.
\end{equation}
The relative breaking $b_{exc}$ of U-spin can be estimated, 
e.g. from spectroscopy. This leads us to 
\begin{equation}
|b_{exc}| =  \frac{M_{K^*} - m_\rho}{\frac{1}{2}(M_{K^*} + m_\rho)}
= 14 \%. 
\end{equation}
Certainly, other estimates are also possible, such as a comparison of
$f_\rho$ and $f_{K^*}$. In this case one finds 
a very small U-spin breaking. Using the 
more conservative value for
$b_{exc}$, which is also compatible with sum rule calculations of form factors 
(see  \cite{Ali:1994vd}), one arrives at
the standard-model prediction for the
difference of branching ratios
\begin{equation} \label{resexc1}
|\Delta {\cal B}(B^- \to K^{*-} \gamma) +
\Delta {\cal B}(B^- \to \rho^- \gamma)| \sim 4 \times 10^{-8}
\end{equation}
Note that the right-hand side is model-dependent.  Still
(\ref{resexc1}) is of some use, since a value
significantly above this estimate would be a strong hint that non-CKM
sources of CP violation are active.

The issue is more attractive in the inclusive modes.
Due to the  $1/m_b$ expansion for the inclusive process,
 the leading contribution is
the free $b$-quark decay. In particular, there is no sensitivity to
the spectator quark and thus one arrives at the following relation
for the CP rate asymmetries
\cite{mannelhurth}:
\begin{equation} \label{resincg1}
\Delta \Gamma (B \to X_s \gamma) +
\Delta \Gamma (B \to X_d \gamma) = b_{inc} \Delta_{inc}. 
\end{equation}

In this framework one relies on
parton-hadron duality (besides in the long-distance contribution
from up-quark loops, which is found to be rather small \cite{LDUP}). 
So one can actually compute the breaking 
of U-spin by keeping a nonvanishing strange quark mass. 
However, it is a formidable task to do this for the CP asymmetries
and it has not yet been done. 
The typical size of $b_{inc}$ can be roughly 
estimated to be of the order of 
$|b_{inc}| \sim m_s^2/m_b^2 \sim 5 \times 10^{-4}$;  
$|\Delta_{inc}|$  is again the average of the moduli of the two CP rate
asymmetries. These have been calculated (for vanishing strange quark mass),
e.g. in  \cite{AG7} and thus one arrives at 
\begin{equation} \label{resinc3}
| \Delta {\cal B}(B \to X_s \gamma) +
\Delta {\cal B}(B \to X_d \gamma) | \sim 1 \cdot 10^{-9}. 
\end{equation}
Again, any measured value in significant deviation of (\ref{resinc3}) 
would be an indication of new sources of CP violation. Although 
only an estimate is given here, it should again  be stressed  
that in the inclusive mode the
right-hand side in (\ref{resinc3})  can be computed in a
model-independent way with the help  of the heavy mass expansion.

\setcounter{equation}{0}
\section{$K_L \rightarrow \pi^0\nu\bar{\nu}$ and $K^+ \rightarrow \pi^+ \nu 
\bar{\nu}$} 
\label{kaonsection}

\begin{figure}
\begin{center}
\epsfig{figure=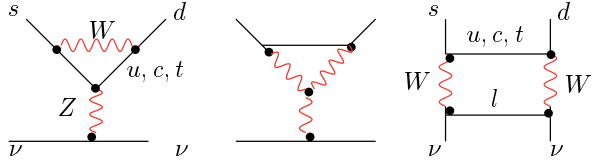,width=8cm}
\end{center}
\vspace{-0.3cm}
\caption{One-loop diagrams contributing to $K \to\pi \nu\bar\nu$.}
\label{SMkaon}
\end{figure}

The rare decays $K_L\to\pi^0\nu\bar\nu$ and $K^+\to\pi^+\nu\bar\nu$ 
represent complementary opportunities for precision flavour physics.
They are flavour changing current processes induced at the one-loop
level (see fig. \ref{SMkaon}) and are exceptionally clean processes.
In particular, the $K_L\to\pi^0\nu\bar\nu$ amplitude can be calculated
with a theoretical uncertainty below $3 \%$ \cite{Buras93}.

This implies the important role of these decay modes for 
CKM phenomenology:
they play a unique role among $K$ decays, like  
the $B_d\to\psi K_S$ mode among 
the $B$ decays.
They allow a measurement of one angle of the unitarity triangle
without  any hadronic  uncertainties to a precision 
comparable to that obtained by 
the $B_d\to\psi K_S$ mode before the LHC era \cite{Buchalla:1994}:
\begin{equation}
\label{sin2beta}
\sin 2\beta=\frac{2 r_s}{1+r^2_s} 
\qquad r_s=\sqrt{\sigma}{\sqrt{\sigma(B_1-B_2)}-P_0(K^+)\o\sqrt{B_2}} 
\end{equation}
where $\sigma$ is just related to the Wolfenstein parameter
$\lambda=0.22$ via 
$(1-\lambda^2/2)^{-2}$; $P_0(K^+)=0.40\pm 0.06$ is the internal charm 
contribution to  $K^+\to\pi^+\nu\bar\nu$; this quantity is known
up to next-to-leading QCD precision, and the dependence on $V_{tb}$ 
is only of second order in $\lambda$; 
$B_1$ and $B_2$ represent here the reduced
branching ratios $ B_1= B(\kpn) / (4.11\times 10^{-11})$ and 
$B_2= B(\klpn) /  (1.80 \times 10^{-10})$.

The time-integrated CP violating asymmetry
in $B^0_d\to\psi K_S$ is given by
$A_{CP}(\psi K_S)=-\sin 2\beta x_d / (1+x^2_d)$
where $x_d=\Delta m/\Gamma$ gives the size of $B^0_d-\bar B^0_d$
mixing. With $(sin 2\beta)_{\pi\nu\bar{\nu}} = 
(sin 2\beta)_{\psi K_s}$,  one obtains an
interesting connection between rare $K$ decays and $B$ physics:
\begin{equation}
\label{golden}
{2 r_s(B_1,B_2)\o 1+r^2_s(B_1,B_2)}=-A_{CP}(\psi K_S){1+x^2_d\o x_d}, 
\end{equation}  
which must be satisfied in the SM. As was stressed 
in~\cite{Buchalla:1994},  
all quantities in this `golden relation'~(\ref{golden}) - 
except for $P_0(K^+)$ - 
can be directly measured experimentally and the relation is
almost independent of $V_{cb}$.

Besides their  rich CKM phenomenology,  the decays  
$K_L \rightarrow \pi^0\nu\bar{\nu}$ and 
$K^+ \rightarrow \pi^+ \nu \bar{\nu}$ 
 are  very sensitive to new physics beyond the SM.
In addition, the theoretical information is very clean
and, thus, the measurement of these decays leads  to 
very accurate constraints on any 
new physics model. 
Moreover, there is the possibility that these clean rare decay modes
themselves lead to first evidence of new physics when the measured 
decay rates are not compatible with the SM.

\begin{figure}
\begin{center}
\epsfig{figure=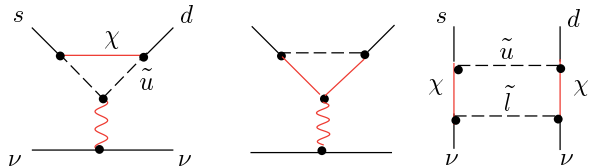,width=8cm}
\end{center}
\vspace{-0.3cm}
\caption{Supersymmetric contributions to $K \to\pi \nu\bar\nu$.}
\label{SUSYkaon}
\end{figure}

New physics contributions in  
$K_L \rightarrow \pi^0\nu\bar{\nu}$ and 
$K^+ \rightarrow \pi^+ \nu \bar{\nu}$ can be parametrized 
in a model-independent way by two parameters
which quantify the violation of the golden 
relation~(\ref{golden})~\cite{Nir98,Buras:1998}.
New effects in supersymmetric models can get induced
through new box diagram and penguin diagram contributions involving
new particles such as charged Higgs or  charginos and stops 
(fig.~\ref{SUSYkaon}), 
replacing the $W$ boson and the up-type quark 
of the SM (fig.~\ref{SMkaon}).

In the constrained minimal supersymmetric standard model 
(MSSM), where all flavour changing effects 
are induced by contributions proportional to the  CKM mixing 
angles the golden relation~(\ref{golden}) is valid. Thus, the 
measurements of ${\cal B}(K_L \rightarrow \pi^0\nu\bar{\nu}$) and 
${\cal B}(K^+ \rightarrow \pi^+ \nu \bar{\nu}$) still directly 
determine the angle $\beta$, and 
a sigificant violation  of (\ref{golden})  
would rule out this model.

At the present experimental status of supersymmetry,
however,  a model-independent analysis including also flavour change
through the squark mass matrices is more suitable. 
If the  new sources of flavour change get parametrized by 
the mass-insertion approximation, an expansion 
of the squark mass matrices around their diagonal, 
it turns out that SUSY contributions in this more general setting
of the unconstrained MSSM 
allow  for a  significant violation of the golden rule. 
An enhancement of the branching ratios by an order of magnitude  
(in the case of $K^+ \rightarrow \pi^+ \nu \bar{\nu}$ by a factor
3) compared with the SM values is possible,
mostly due to the chargino-induced Z-penguin 
contribution~\cite{Colangelo:1998}.
Recent analyses~\cite{Colangelo:1998,Burasb,all}
within the uMSSM
focused on the  correlation of rare decays and $\epsilon' / \epsilon$,
and led to  reasonable upper bounds for
the branching ratios: 
${\cal B}(K_L \rightarrow \pi^0\nu\bar{\nu}) \leq  1.2 \times 10^{-10}$,
and ${\cal B}(K^+ \rightarrow \pi^+ \nu \bar{\nu}) \leq 1.7 \times 10^{-10}$.
which should be compared with the latest numerical SM predictions
\cite{Buras999}:  
${\cal B}(K^+\to\pi^+\nu\bar\nu) = (7.9 \pm 3.1) \times 10^{-11}$, 
${\cal B}(K_L\to\pi^0\nu\bar\nu) = (2.8\pm 1.1) \times 10^{-11}$.

The rare decays $K^+ \rightarrow \pi^+ \nu \bar{\nu}$ and 
 $K_L \rightarrow \pi^0 \nu \bar{\nu}$ are  
specifically interesting in view of the suggested  experiments 
at the Brookhaven 
laboratory (USA) \cite{brookhavennew}  and at FERMILAB (USA) \cite{fermilabnew}
\cite{fermilabnew2}
 and at  KEK (Japan) \cite{keknew}.
The current Brookhaven experiment E787 has already 
observed a single, but clean 
candidate event for $K^+\to\pi^+\nu\bar\nu$ in 1997 which 
corrsponds to the following branching ratio \cite{brookhavenold}:
\begin{equation}
{\cal B}(K^+\to\pi^+\nu\bar\nu)=(1.5^{+3.4}_{-1.2})\times 10^{-10}.
\end{equation}
For the $K_L\to\pi^0\nu\bar\nu$ mode, there is only an upper
bound available \cite{ktevold}:
\begin{equation}
{\cal B}(K_L\to\pi^0\nu\bar\nu) < 5.9 \times 10^{-7}.
\end{equation}
An indirect upper bound on ${\cal B}(K_L\to\pi^0\nu\bar\nu)$, 
using the current limit on 
${\cal B}(K^+\to\pi^+\nu\bar\nu)$ and isospin symmetry, can be
placed \cite{Grossman98} at $2 \times 10^{-9}$.

\section{Summary}
In this paper we have reviewed the status of inclusive rare $B$ decays,
highlighting recent developments. 
These  decays  give special insight into the CKM matrix; 
moreover,  as flavour changing neutral current processes,  they are
loop-induced and therefore  particularly sensitive to new physics.

Decays modes such as  
$B \rightarrow X_s \gamma$ and  
$B \rightarrow X_s l^+l^-$  (with specific kinematic cuts)
 represent laboratories for 
perturbative QCD.
Nonperturbative contributions play a subdominant role and they  
are under control thanks to the heavy mass expansion. 
The inclusive rare $B$  decays  are or will be accessible at the present 
\mbox{$e^+$ $e^-$} machines with their low background 
 and their kinematic constraints (CLEO, BaBar, BELLE) and will make 
precision flavour physics possible
in the near future.  

Significant  progress has been made during the last couple of years.
The calculation of NLL (or even NNLL)  QCD corrections to these decay modes 
has been performed. The theoretical uncertainty has been reduced below 
the $10 \%$ level. 
As was emphasized, the step from LL to NLL precision within the 
framework  of the 
renormalization group improved perturbation theory is not only a
quantitative,
but also a qualitative one, which tests the validity of the perturbative
approach in the given problem.

Inclusive rare $B$ decays allow for an indirect search for new physics,
a strategy complementary to the direct production of new (supersymmetric) 
particles,  which is reserved for the planned 
hadronic machines such as the LHC at CERN. 
However, the indirect search at the 
$B$ factories  will imply  significant restrictions for 
the parameter space of supersymmetric models and will thus lead to important 
theoretically clean information  for the direct search of 
supersymmetric particles.
Within supersymmetric models the QCD calculation of the inclusive 
rare $B$ decays has not reached
the sophistication of the corresponding SM calculation. However,
NLL analyses in specific scenarios already show that bounds on the 
parameter space of nonstandard models are rather sensitive to NLL
QCD contributions. 

Detailed measurements of CP asymmetries in rare $B$ decays will also 
be possible
in the near future. They will allow for a stringent and clean test if the CKM
matrix is indeed the only source of CP violation. 

The rare kaon decays, 
$K^+ \rightarrow \pi^+ \nu \bar{\nu}$ and 
 $K_L \rightarrow \pi^0 \nu \bar{\nu}$,  
offer complementary opportunities for precision flavour   
physics. Besides the current Brookhaven experiment, several 
more are planned or suggested to explore these 
theoretically clean decay modes.

\section*{Acknowledgements}
I am grateful to Gerhard Buchalla and Christoph Greub for careful
reading of the manuscript.
I thank Thomas Besmer, Francesca Borzumati, Christoph Greub, Thomas Mannel,
Mikolaj Misiak and Daniel Wyler for pleasant collaborations and countless
discussions on many issues reviewed
in this paper.

\end{document}